\documentclass[onecolumn,superscriptaddress,preprintnumbers,amsmath,amssymb,prc]{revtex4}
\usepackage[capitalize]{cleveref}
\usepackage{graphicx}
\usepackage{subfigure}
\usepackage{dcolumn}
\usepackage{bm}
\usepackage{color}

\begin{document}
\title{Impact of magnetic-field fluctuations on measurements of the chiral magnetic effect in collisions of isobaric nuclei}
\author{Xin-Li Zhao}
\affiliation{Shanghai Institute of Applied Physics, Chinese Academy of Sciences, Shanghai 201800, China}
\affiliation{University of Chinese Academy of Sciences, Beijing 100049, China}
\author{Guo-Liang Ma}
\email[]{glma@fudan.edu.cn}
\affiliation{Key Laboratory of Nuclear Physics and Ion-beam Application (MOE), Institute of Modern Physics, Fudan University, Shanghai 200433, China}
\affiliation{Shanghai Institute of Applied Physics, Chinese Academy of Sciences, Shanghai 201800, China}
\author{Yu-Gang Ma}
\email[]{mayugang@fudan.edu.cn}
\affiliation{Key Laboratory of Nuclear Physics and Ion-beam Application (MOE), Institute of Modern Physics, Fudan University, Shanghai 200433, China}
\affiliation{Shanghai Institute of Applied Physics, Chinese Academy of Sciences, Shanghai 201800, China}


\begin{abstract}

We investigate the properties of electromagnetic fields in isobaric $_{44}^{96}\textrm{Ru}+\,_{44}^{96}\textrm{Ru}$ and $_{40}^{96}\textrm{Zr}+\,_{40}^{96}\textrm{Zr}$ collisions at $\sqrt{s}$ = 200 GeV by using a multiphase transport model, with special emphasis on the correlation between magnetic field direction and participant plane angle $\Psi_{2}$ (or spectator plane angle $\Psi_{2}^{\rm SP}$), i.e. $\langle{\rm cos}\ 2(\Psi_B - \Psi_{2})\rangle$ [or $\langle{\rm cos}\ 2(\Psi_B - \Psi_{2}^{\rm SP})\rangle$]. We  confirm that the magnetic fields of $_{44}^{96}\textrm{Ru}+\,_{44}^{96}\textrm{Ru}$ collisions are stronger than those of $_{40}^{96}\textrm{Zr}+\,_{40}^{96}\textrm{Zr}$ collisions due to their larger proton fraction. We find that the deformation of nuclei has a non-negligible effect on $\langle{\rm cos}\ 2(\Psi_B - \Psi_{2})\rangle$ especially in peripheral events. Because the magnetic-field direction is more strongly correlated with $\Psi_{2}^{\rm SP}$ than with $\Psi_{2}$, the relative difference of the chiral magnetic effect observable with respect to $\Psi_{2}^{\rm SP}$ is expected to be able to reflect much cleaner information about the chiral magnetic effect with less influences of deformation.

\end{abstract}


\maketitle

\section{Introduction}
\label{introduction}

Lattice QCD calculations predicted that quarks and gluons are deconfined with their partonic degrees of freedom under the condition of high temperatures or the high baryon chemical potential, i.e., the formation of quark-gluon plasma (QGP). Relativistic heavy ion collisions are believed to be able to reach the condition of creating the QGP. On the other hand, a nonzero axial charge density of the QGP with a large magnetic-field ${\bf B}$ can lead to a dipole charge separation along the ${\bf B}$ direction, i.e., the so-called chiral magnetic effect (CME), which results in a generation of a vector current {\bf J}~\cite{Kharzeev:2015znc,Fukushima:2008xe,Kharzeev:2004ey,Kharzeev:2007tn,Hattori:2016emy},
\begin{eqnarray}
{\bf J}=\sigma _{5}{\bf B},\qquad\sigma _{5}=\frac{Qe}{2\pi ^{2}}\mu _{5},
\label{current}
\end{eqnarray}%
where $\sigma_{5}$ is the chiral magnetic conductivity and $\mu_{5}$ is the chiral chemical potential arising from the nonzero axial charge density.

To measure the CME signal, people usually measure charge azimuthal correlation~\cite{Abelev:2009ac,Abelev:2009ad,Abelev:2012pa} between two particles $\alpha$ and $\beta$, which is defined as
\begin{eqnarray}
\gamma=\langle\rm cos(\phi_{\alpha}+\phi_{\beta} - 2\Psi_{\rm RP})\rangle,
\label{gamma}
\end{eqnarray}%
where $\phi_{\alpha}$ and $\phi_{\beta}$ are the azimuthal angles of two charged particles and $\Psi_{\rm RP}$ is the reaction plane angle, which is usually represented by the second order of the event participant plane $\Psi_{2}$. From the CME expectation, the charge azimuthal correlation $\Delta \gamma = \gamma_{\rm opp} - \gamma_{\rm same}$ (the difference between opposite-pair and same-pair correlations) is expected to be proportional to $B^{2}$ and ${\rm cos}\ 2(\Psi_B - \Psi_{2})$ ~\cite{Bloczynski:2012en,Deng:2016knn}, i.e.,
\begin{eqnarray}
\Delta \gamma \propto \langle B^{2}{\rm cos}\ 2(\Psi_B - \Psi_{2})\rangle.
\label{delta}
\end{eqnarray}%

However, the current main difficulty of measuring the CME signal is some backgrounds which we do not understand clearly~\cite{Bzdak:2010fd,Schlichting:2010qia,Wang:2009kd,Ma:2011uma}. For example, one of the difficulties of the CME observable interpretation is due to a large part of background contribution stemming from the coupling of resonance decay correlations and the flow $v_{2}$ arising from participant geometry ~\cite{Zhao:2018blc,Wang:2016iov,Adamczyk:2013kcb}. To isolate the influence of those backgrounds, the isobar program at the Relativistic Heavy Ion Collider (RHIC) has been proposed and it collides $_{44}^{96}\textrm{Ru}+\,_{44}^{96}\textrm{Ru}$ and $_{40}^{96}\textrm{Zr}+\,_{40}^{96}\textrm{Zr}$ elements since they have a same nucleon number but the 10$\%$ difference in proton number. The same nucleon number indicates they should have similar bulk backgrounds (e.g., flow), however, the different proton number means they carry different magnitudes of magnetic fields. Therefore, the CME signal (due to the CME current ${\bf J}$ ) is expected to be different between the two isobaric collisions, as illustrated by Eq.~(\ref{current}). There has been some interesting research on isobaric collisions, see Refs.~\cite{Voloshin:2010ut,Deng:2016knn,Shi:2017ucn,Magdy:2018lwk,Huang:2017azw,Deng:2018dut,Xu:2017zcn,Li:2018oec,Sun:2018idn}.

If there are similar or even the same backgrounds in two isobaric collisions, the difference of the CME observable between two isobaric collisions is expected to be mainly due to the differences from the squared magnetic field and the correlation between magnetic-field direction $\Psi_{B}$ and participant plane $\Psi_{2}$ from Eq.~(\ref{delta}). Meanwhile, because the magnetic field is mainly induced by spectator protons, people also proposed to replace the participant plane $\Psi_{2}$ with the spectator plane $\Psi_{2}^{\rm SP}$, which is believed to be more strongly correlated with $\Psi_{B}$~\cite{Xu:2017zcn,Voloshin:2018qsm}. In this paper, we focus on not only the magnetic field, but also the two correlations between magnetic-field direction $\Psi_{B}$ and participant plane angle $\Psi_{2}$ and between $\Psi_{B}$ and the spectator plane angle $\Psi_{2}^{\rm SP}$. We systematically study $_{44}^{96}\textrm{Ru}+\,_{44}^{96}\textrm{Ru}$ collisions and $_{40}^{96}\textrm{Zr}+\,_{40}^{96}\textrm{Zr}$ collisions by a multiphase transport model (AMPT) model. Based on the above, the implications of our results to the future CME analysis in the isobaric experiment will be discussed.

The paper is organized as follows. In Sec.~\ref{GS}, we provide a brief introduction to the AMPT model, our isobaric deformation settings, and the method to calculate magnetic fields. The numerical results for the properties of electromagnetic fields and some related correlations are presented and discussed in detail in Sec.~\ref{results}. Section~\ref{summary} contains our conclusions.

\section{GENERAL SETUP}
\label{GS}
\subsection{AMPT model}

In this paper, we take advantage of a AMPT model~\cite{Lin:2004en} to investigate isobaric collisions. There are two versions of the AMPT model, the default version and the version with a string-melting mechanism. Both versions contain four important evolution stages of heavy ion collisions: initial state, parton cascade, hadronization, and hadron rescatterings. They both use the $\rm \tiny{HIJING}$ model~\cite{Wang:1991hta,Gyulassy:1994ew} for generating the initial state of collisions. The main difference between the two versions is that in the string-melting version, strings and minijets are melted into partons so that there are more partons participating in the parton cascade than the default version. Therefore, the string-melting version can better describe the cases when the QGP is produced, such as heavy ion collisions at the RHIC and Large Hadron Collider energies. The string-melting version currently only considers elastic collision processes between two partons~\cite{Zhang:1997ej}, hadronization is simulated by a simple quark combination model, and hadron rescatterings are described by a hadron transport model~\cite{Li:1995pra}. In this paper, we choose the string-melting version to simulate $_{44}^{96}\textrm{Ru}+\,_{44}^{96}\textrm{Ru}$ and $_{40}^{96}\textrm{Zr}+\,_{40}^{96}\textrm{Zr}$ collisions at the top RHIC energy of $\sqrt{s}$ = 200 GeV. In our convention, we choose the $x$ axis along the direction of impact parameter $b$ from the target center to the projectile center, the $z$ axis along the beam direction, and the $y$ axis perpendicular to the $x$ and $z$ directions.

\subsection{Geometry configuration of isobaric collisions}

For modeling $_{44}^{96}\textrm{Ru}$ and $_{40}^{96}\textrm{Zr}$ in the $\rm \tiny{HIJING}$ model, the spatial distribution of nucleons in their rest frame can be written in the Woods-Saxon form (in spherical coordinates),
\begin{eqnarray}
\rho (r,\theta )=\rho _{0}/(1+{\rm exp}((r-R_{0}-\beta _{2}R_{0}Y_{2}^{0}(\theta))/a)),
\label{rho}
\end{eqnarray}%
where the normal nuclear density $\rho _{0} = 0.16\ {\rm fm}^{-3},\ R_{0}$ is the radius of nucleus ( $R_{0}$ = 5.085 fm for $_{44}^{96}\textrm{Ru}$ and $R_{0}$ = 5.02 fm for $_{40}^{96}\textrm{Zr}$), $a$ is the surface diffuseness parameter, and $\beta _{2}$ is the deformity of nucleus. For $_{44}^{96}\textrm{Ru}$ and $_{40}^{96}\textrm{Zr}$, the parameter $a$ is almost identical, $a \approx$ 0.46 fm. At present, we can not confirm the $\beta_{2}$ of $_{44}^{96}\textrm{Ru}$ and $_{40}^{96}\textrm{Zr}$ because there are two cases of $\beta_{2}$~\cite{Shou:2014eya} from $e-A$ scattering experiments~\cite{Raman:1201zz,Pritychenko:2013gwa} and comprehensive model deductions~\cite{Moller:1993ed}. For the first case (denoted as case 1 thereafter), $_{44}^{96}\textrm{Ru}$ is more deformed than $_{40}^{96}\textrm{Zr}$, i.e., $\beta_{2}^{Ru}$ =0.158 and $\beta _{2}^{Zr}$=0.08. However, the second case (denoted as case 2 thereafter) is the opposite, i.e., $\beta _{2}^{\rm Ru}$ = 0.053 and $\beta _{2}^{\rm Zr}$ = 0.217. As shown in Ref.~\cite{Deng:2016knn}, the systematic uncertainty has little influence on the multiplicity distribution. We focus on its impact on the CME signal of the correlator $\Delta\gamma$. To cancel some theoretical uncertainties~\cite{Deng:2016knn}, we can take the ratio of the relative difference between the two collisions. The definition of the relative ratio in a quantity $Q$ between $_{44}^{96}\textrm{Ru}+\,_{44}^{96}\textrm{Ru}$ and $_{40}^{96}\textrm{Zr}+\,_{40}^{96}\textrm{Zr}$ collisions is
\begin{eqnarray}
R_{Q}\equiv 2(Q^{\rm Ru+Ru}-Q^{\rm Zr+Zr})/(Q^{\rm Ru+Ru}+Q^{\rm Zr+Zr})
\label{ratio}
\end{eqnarray}%
and $Q$ can represent $\langle e|B|/m_{\pi}^{2}\rangle$, $\langle{\rm cos}\ 2(\Psi_B - \Psi_{2})\rangle$, $\langle{\rm cos}\ 2(\Psi_B - \Psi_{2}^{\rm SP})\rangle$, $\langle(eB/m_{\pi}^{2})^{2}{\rm cos}\ 2(\Psi_B - \Psi_{2})\rangle$ and $\langle(eB/m_{\pi}^{2})^{2}{\rm cos}\ 2(\Psi_B - \Psi_{2}^{\rm SP})\rangle$ in our calculations. If $R_{Q}$ is close to zero, it implies a similarity between two isobaric systems, however, it implies a big difference if  $R_{Q}$ is far away from zero. For relative differences of deformation, $R_{\beta _{2}}$ = 0.33 for case 1, but $R_{\beta _{2}} = -1.43$ for case 2, which implies a larger deformation difference for case 2 than that for case 1.

\subsection{Calculations of the electromagnetic field}

Following Refs.~\cite{Bzdak:2011yy,Deng:2012pc,Zhao:2017rpf,Deng:2017ljz}, we use the same way to calculate the initial electromagnetic fields as
\begin{eqnarray}
e{\bf E}(t,{\bf r})&=&\frac{e^2}{4\pi}{\sum\limits_{n}}Z_{n} \frac{{\bf R}_n- R_n{\bf v}_n}{(R_n-{\bf R}_n \cdot {\bf v}_n)^3}(1-v_n^2),
\label{elec}\\
e{\bf B}(t,{\bf r})&=&\frac{e^2}{4\pi}{\sum\limits_{n}}Z_{n} \frac{{\bf v}_n \times {\bf R}_n}{(R_n-{\bf R}_n \cdot {\bf v}_n)^3}(1-v_n^2),
\label{magn}
\end{eqnarray}%
where we use natural unit  $\hbar = c = 1,\ Z_n$ is the charge number of the $n$th particle, for the proton it is one, ${\bf R}_n = {\bf r} - {\bf r}_n$ is the relative position of the field point {\bf r} to the source point ${\bf r}_n$, and ${\bf r}_n$ is the location of the $n$th particle with velocity ${\bf v}_n$ at the retarded time $t_{n} = t - |{\bf r} - {\bf r}_n|$. The summations run over all charged protons in the system. For $_{44}^{96}\textrm{Ru}+\,_{44}^{96}\textrm{Ru}$ collisions and $_{40}^{96}\textrm{Zr}+\,_{40}^{96}\textrm{Zr}$ collisions, we need to emphasize that most of our results about their electromagnetic field are calculated at the field point ${\bf r} = (0,0,0)$ at $t = 0$.

\subsection{Calculations of the participant plane and the spectator plane}

We calculate the participant plane $\Psi_2$ by using the spatial distribution of partons from the string-melting mechanism before the parton cascade process starts. The participant plane can be given by
\begin{eqnarray}
\Psi_{2}&=&\frac{1}{2}[{\rm arctan}\frac{\langle r_{p}^{2}{\rm sin}(2\phi_{p})\rangle}{\langle r_{p}^{2}{\rm cos}(2\phi_{p})\rangle}+\pi ],
\label{psi2}
\end{eqnarray}%
where $r_{p}$ is the displacement of the participating partons from field point ${\bf r}$ = (0, 0, 0) and $\phi_{p}$ is the azimuthal angle of the participating partons on the transverse plane~\cite{Alver:2010gr,Ma:2010dv}.

Following Refs.~\cite{Chatterjee:2014sea,Abelev:2013cva,Zhao:2018blc}, we calculate the spectator plane as
\begin{eqnarray}
\Psi_{2}^{\rm SP}&=&\frac{1}{2}{\rm arctan}\frac{\langle r_{s}^{2}{\rm sin}(2\phi_{s})\rangle}{\langle r_{s}^{2}{\rm cos}(2\phi_{s})\rangle},
\label{psi2}
\end{eqnarray}%
where $r_{s}$ is the displacement of spectator neutrons only from one projectile from field point ${\bf r}$ = (0, 0, 0) and $\phi_{s}$ is the azimuthal angle of spectator neutrons only from one projectile in the transverse plane. We check that our results change little even if we use spectator protons. We choose spectator neutrons because the zero-degree calorimeters at STAR Collaboration~\cite{Adler:2001fq} only can measure neutrons. In the above two formulas, the bracket $\langle\cdot\cdot\cdot\rangle$ mean taking the average over all participating partons or all spectator neutrons of projectile, respectively.

\section{Results and discussions}
\label{results}

\subsection{Spatial distributions of electromagnetic fields in isobaric collisions}
\label{resultsA}

\begin{figure*}[htb]
  \setlength{\abovecaptionskip}{0pt}
  \setlength{\belowcaptionskip}{8pt}\centerline{\includegraphics[scale=0.66]{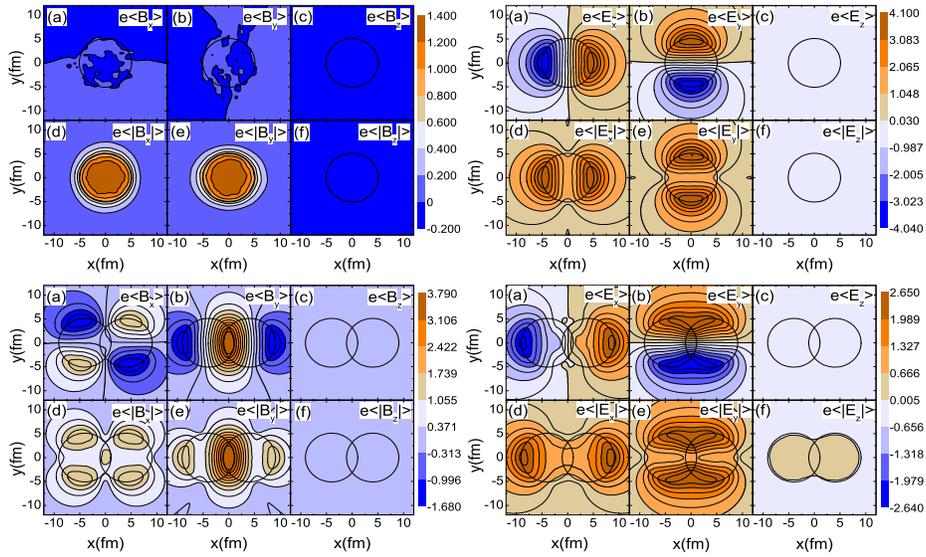}}
\caption{(Color online) The spatial distributions of the electromagnetic fields on the transverse plane at $t = 0$ for $b = 0$ (upper panels) and $b = 8$ fm (lower panels) in $_{44}^{96}\textrm{Ru}+\,_{44}^{96}\textrm{Ru}$ collisions at $\sqrt s = 200$ GeV for case 1 where the unit is $m_\pi^2$. The black solid circles indicate the two colliding nuclei.}\label{fig:spa}
\end{figure*}

Figure~\ref{fig:spa} shows the contour plots of $\langle B_{x, y, z}\rangle$, $\langle|B_{x, y, z}|\rangle$, $\langle E_{x, y, z}\rangle$ and $\langle|E_{x, y, z}|\rangle$ at $t$ = 0 on the transverse plane in $_{44}^{96}\textrm{Ru}+\,_{44}^{96}\textrm{Ru}$ collisions at $\sqrt{s}$ = 200 GeV for case 1 where the  two upper panels are for $b$ = 0 fm and the two lower panels are for $b$ = 8 fm. We find that $\langle|B_{x}|\rangle$ is far less than $\langle|B_{y}|\rangle$ at ${\bf r}={\bf 0}$, what is more, the maximum of the magnetic fields is in field point ${\bf r}={\bf 0}$ for mid central collisions. $\langle E_{x}\rangle$ peaks around $(x$, $y)=(R_{\rm Ru}+b/2$ , $0)$ or $(-R_{\rm Ru}-b/2$ , $0)$, whereas $\langle E_{y}\rangle$ peaks around $(x$, $y)=(0$ , $\pm R_{\rm Ru})$ where $R_{\rm Ru}$ is the radius of the Ru nucleus. Meanwhile, we also study the spatial distributions of electromagnetic fields in $_{44}^{96}\textrm{Ru}+\,_{44}^{96}\textrm{Ru}$ collisions for case 2 and $_{40}^{96}\textrm{Zr}+\,_{40}^{96}\textrm{Zr}$ collisions for case 1 and case 2. We find that their spatial distributions are similar to those in $_{44}^{96}\textrm{Ru}+\,_{44}^{96}\textrm{Ru}$ collisions for case 1. Nevertheless, the fields for $_{40}^{96}\textrm{Zr}+\,_{40}^{96}\textrm{Zr}$ collisions are with smaller magnitudes than those in $_{44}^{96}\textrm{Ru}+\,_{44}^{96}\textrm{Ru}$ collisions due to being with less protons everywhere basically.

\subsection{Centrality dependencies of the electromagnetic fields in isobaric collisions}
\label{resultsB}

\begin{figure*}[htb]
  \begin{minipage}[t]{0.333\linewidth}
    \includegraphics[width=0.95\textwidth]{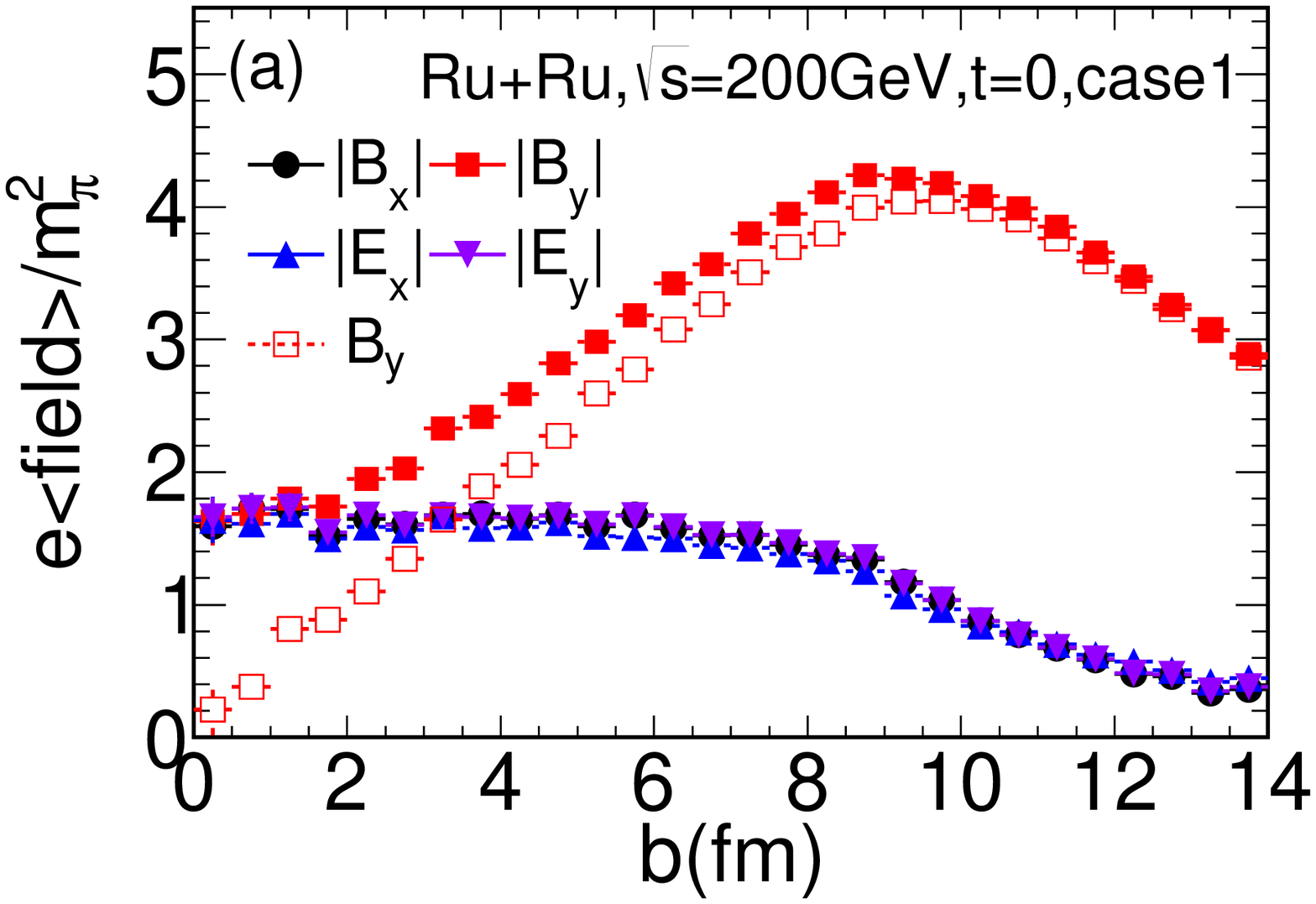}
    \label{fig:side:a}
  \end{minipage}%
  \begin{minipage}[t]{0.333\linewidth}
    \centering
    \includegraphics[width=0.95\textwidth]{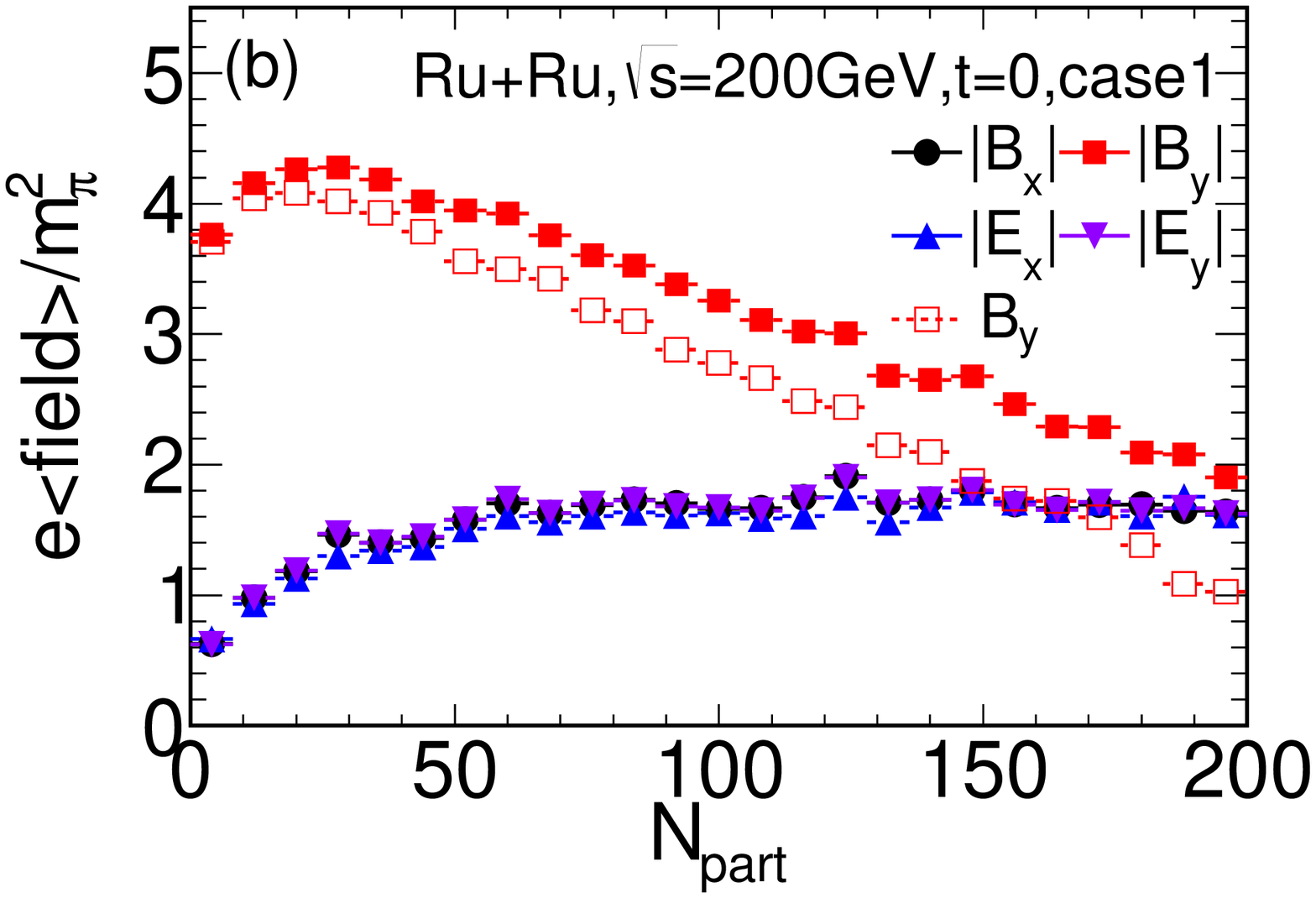}
    \label{fig:side:b}
  \end{minipage}%
  \begin{minipage}[t]{0.333\linewidth}
    \centering
    \includegraphics[width=0.95\textwidth]{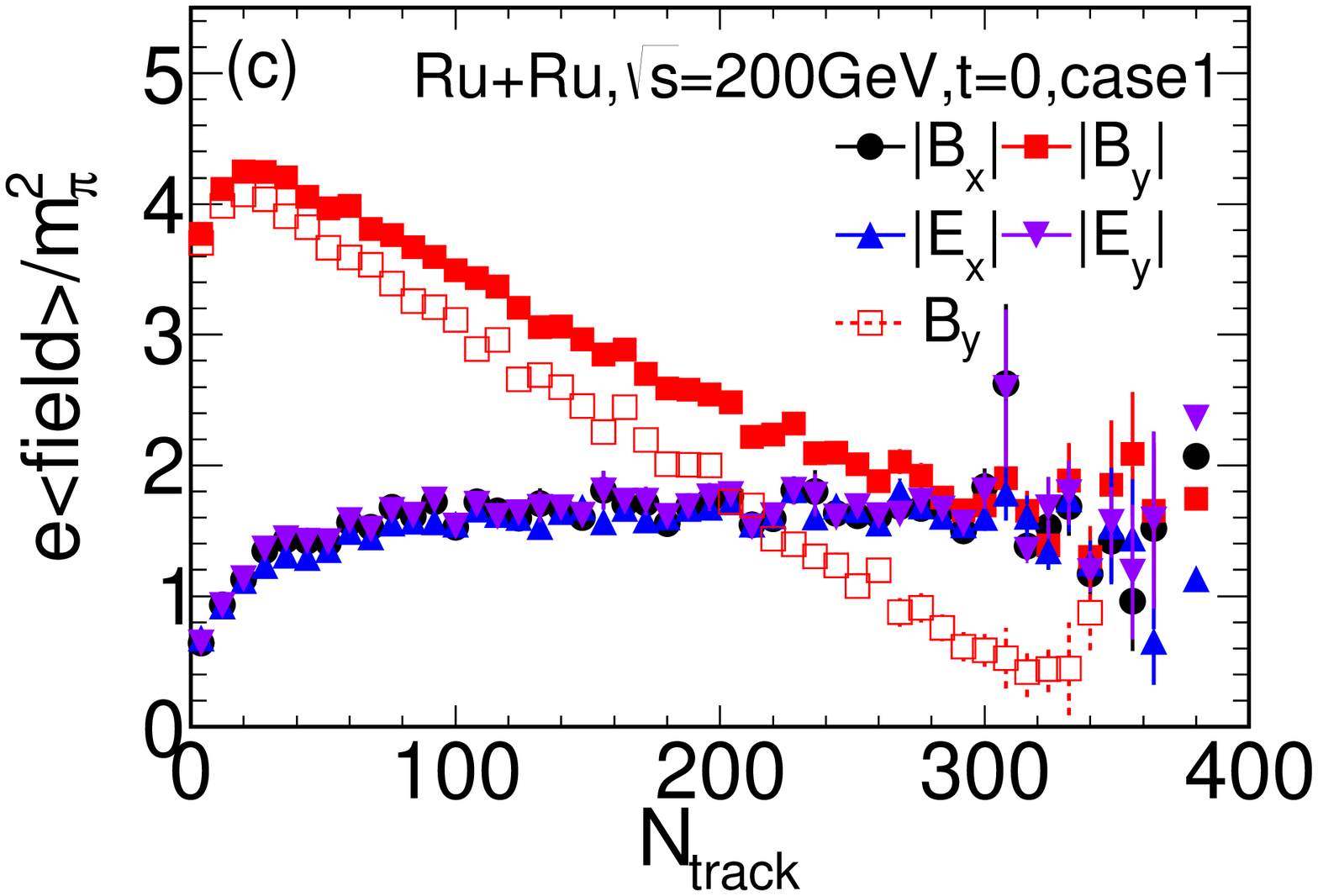}
    \label{fig:side:b}
  \end{minipage}
\caption{(Color online) The electromagnetic fields at $t = 0$ and ${\bf r}={\bf 0}$ as functions of (a) impact parameter $b$, (b) $N_{\rm part}$ and, (c) $N_{\rm track}$ in $_{44}^{96}\textrm{Ru}+\,_{44}^{96}\textrm{Ru}$ collisions at $\sqrt s = 200$ GeV for case 1.}
\label{fig:Ru1EB}
\end{figure*}

\begin{figure*}[htb]
  \begin{minipage}[t]{0.333\linewidth}
    \includegraphics[width=0.95\textwidth]{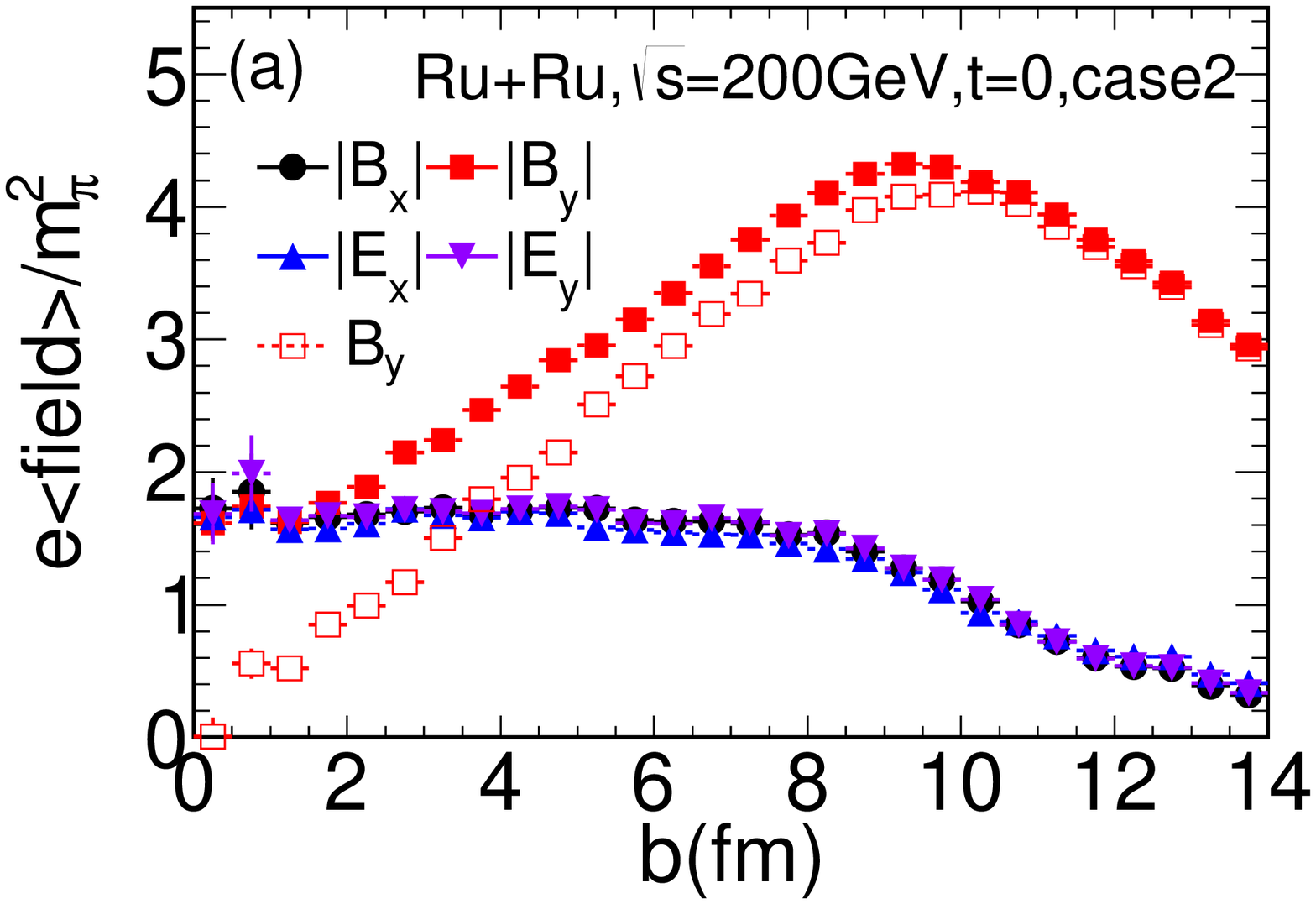}
    \label{fig:side:a}
  \end{minipage}%
  \begin{minipage}[t]{0.333\linewidth}
    \centering
    \includegraphics[width=0.95\textwidth]{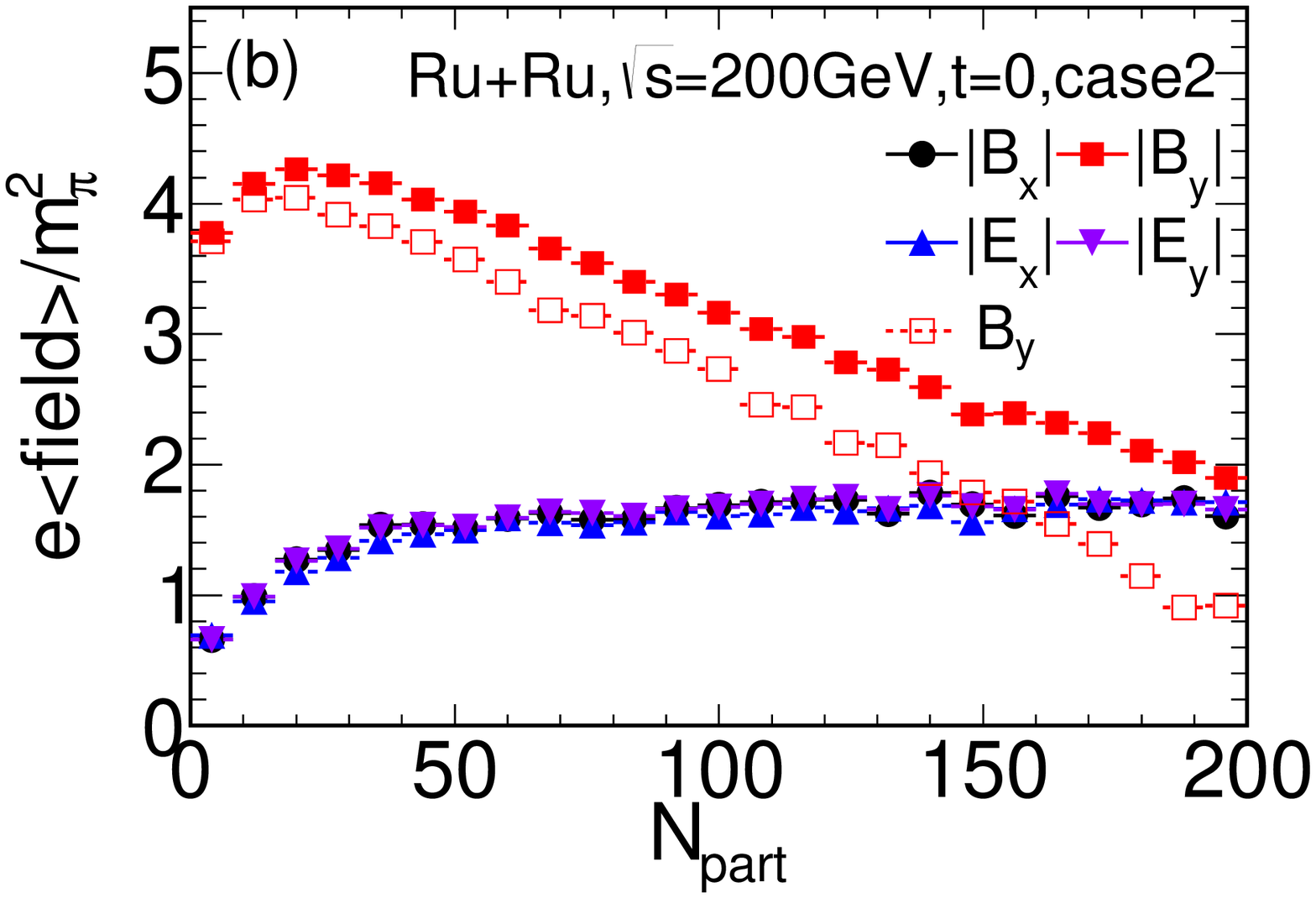}
    \label{fig:side:b}
  \end{minipage}%
  \begin{minipage}[t]{0.333\linewidth}
    \centering
    \includegraphics[width=0.95\textwidth]{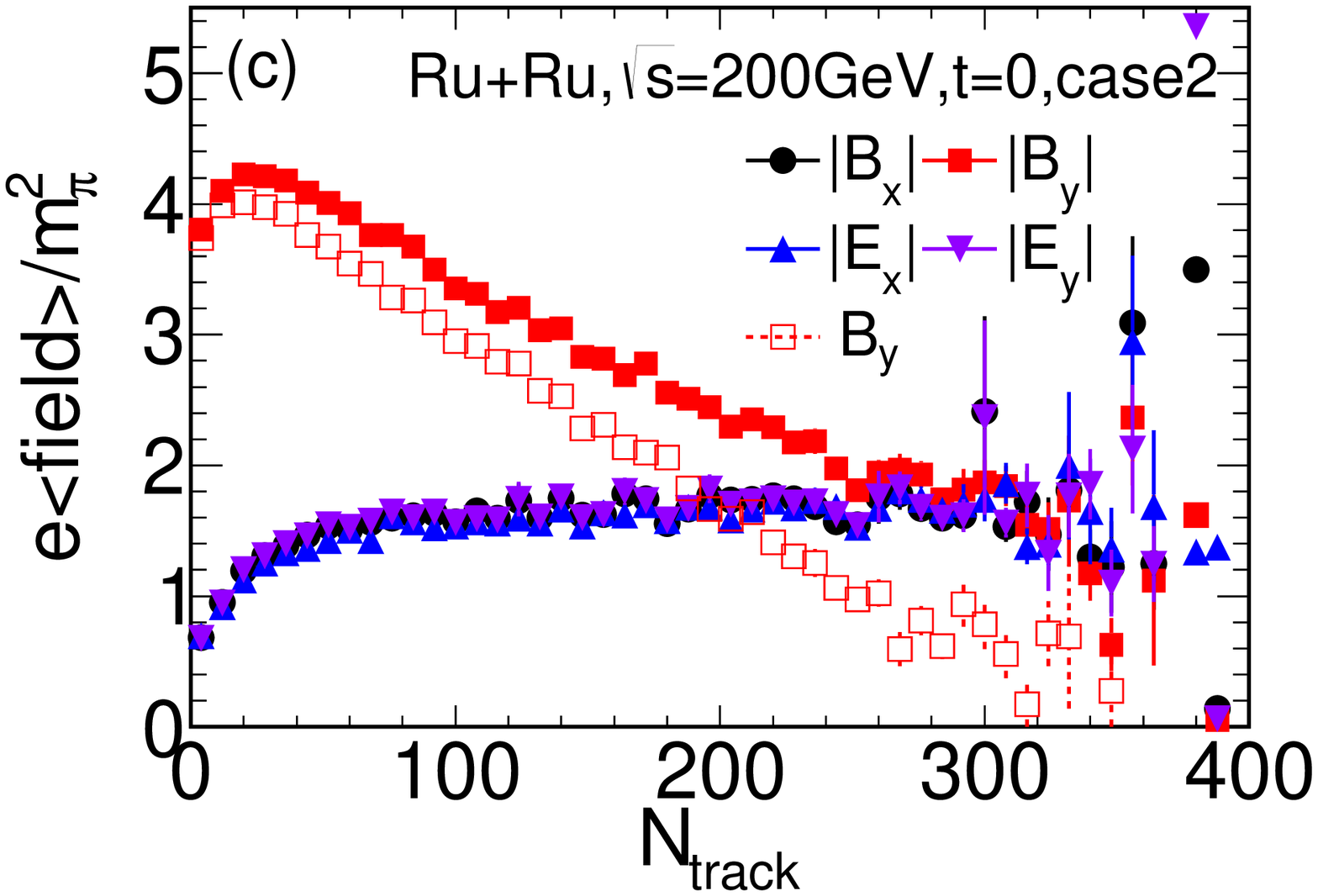}
    \label{fig:side:b}
  \end{minipage}
\caption{(Color online) The electromagnetic fields at $t = 0$ and ${\bf r}={\bf 0}$ as functions of (a) impact parameter $b$, (b) $N_{\rm part}$ and, (c) $N_{\rm track}$ in $_{44}^{96}\textrm{Ru}+\,_{44}^{96}\textrm{Ru}$ collisions at $\sqrt s = 200$ GeV for case 2.}
\label{fig:Ru2EB}
\end{figure*}

\begin{figure*}[htb]
  \begin{minipage}[t]{0.333\linewidth}
    \includegraphics[width=0.95\textwidth]{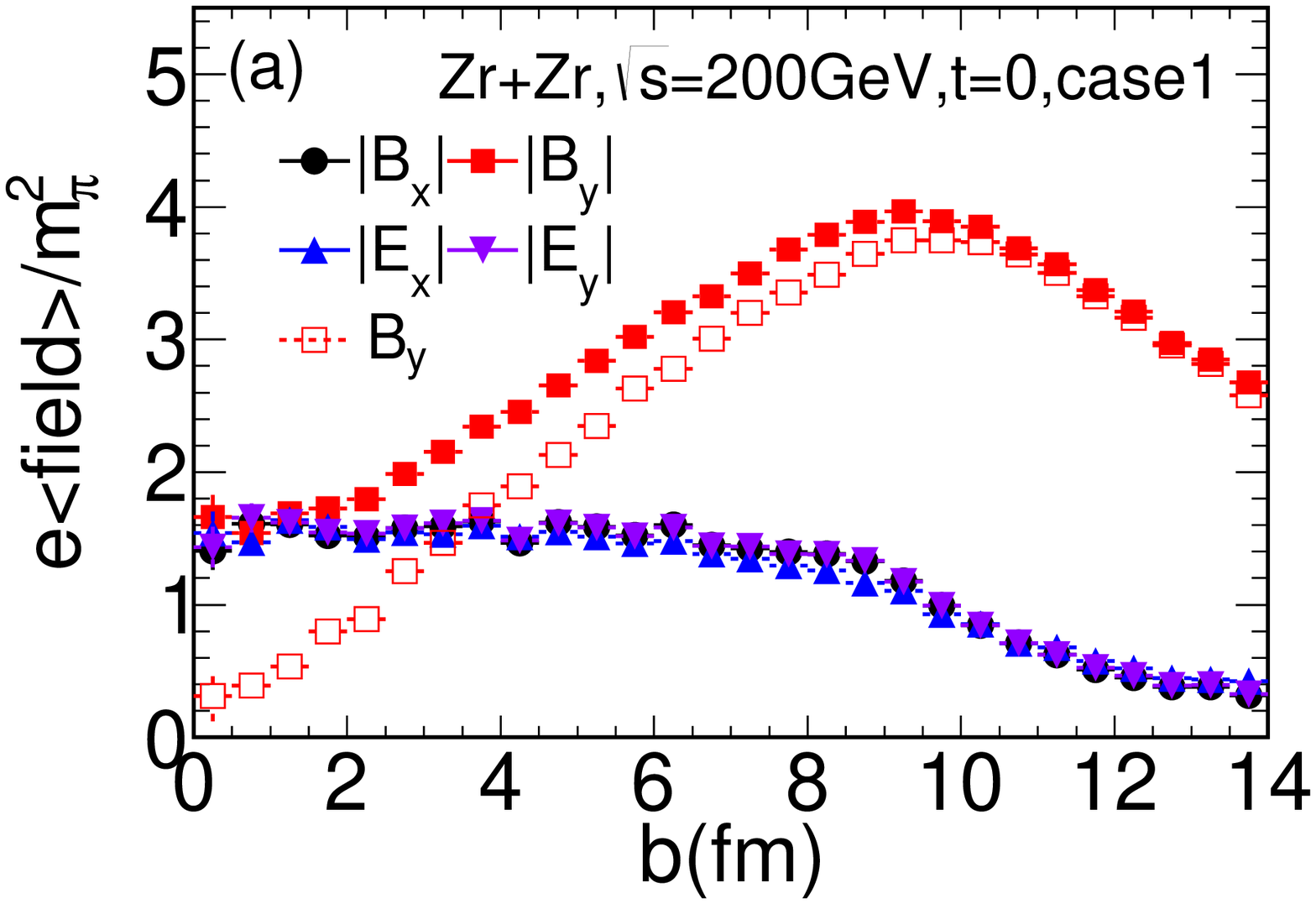}
    \label{fig:side:a}
  \end{minipage}%
  \begin{minipage}[t]{0.333\linewidth}
    \centering
    \includegraphics[width=0.95\textwidth]{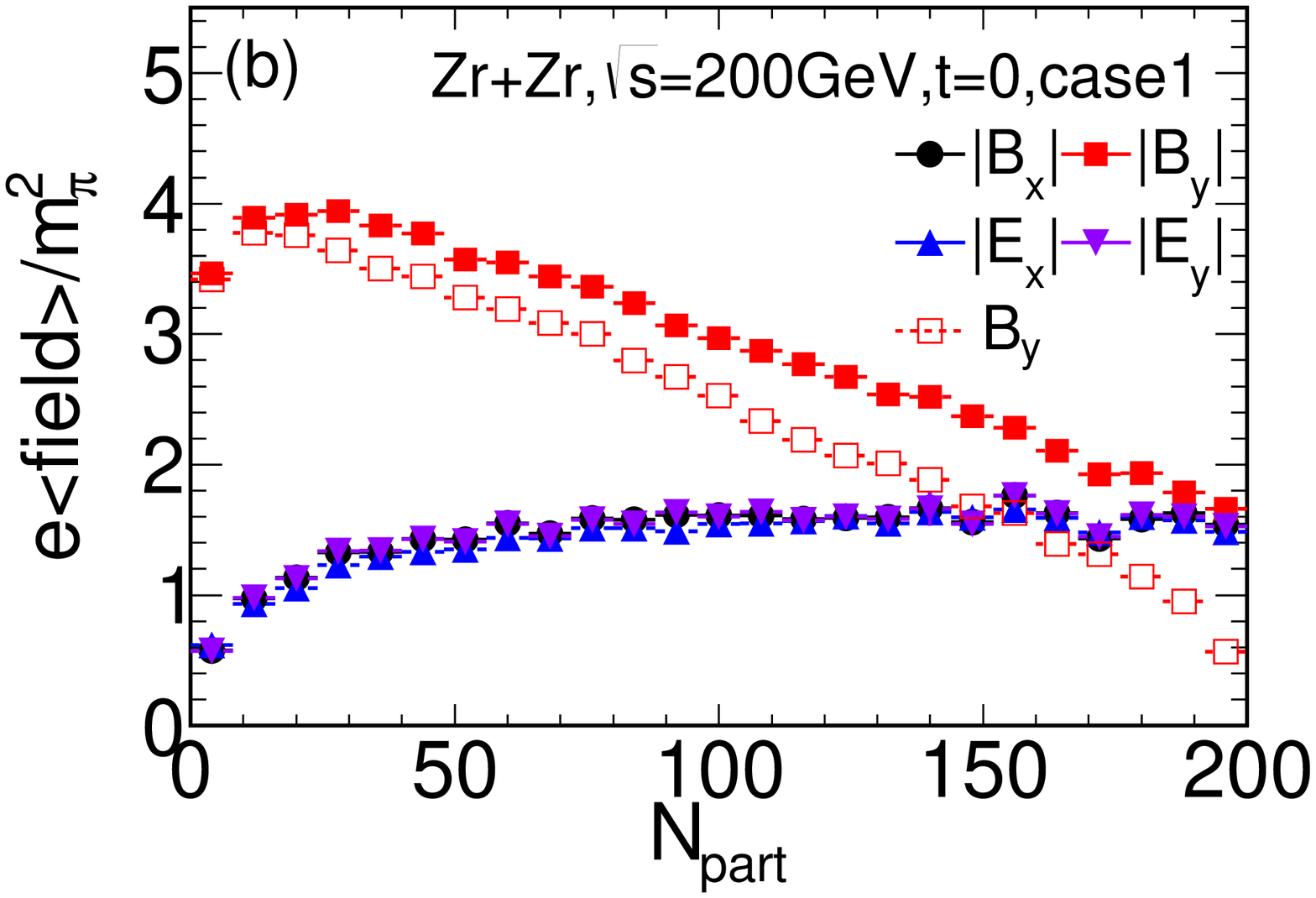}
    \label{fig:side:b}
  \end{minipage}%
  \begin{minipage}[t]{0.333\linewidth}
    \centering
    \includegraphics[width=0.95\textwidth]{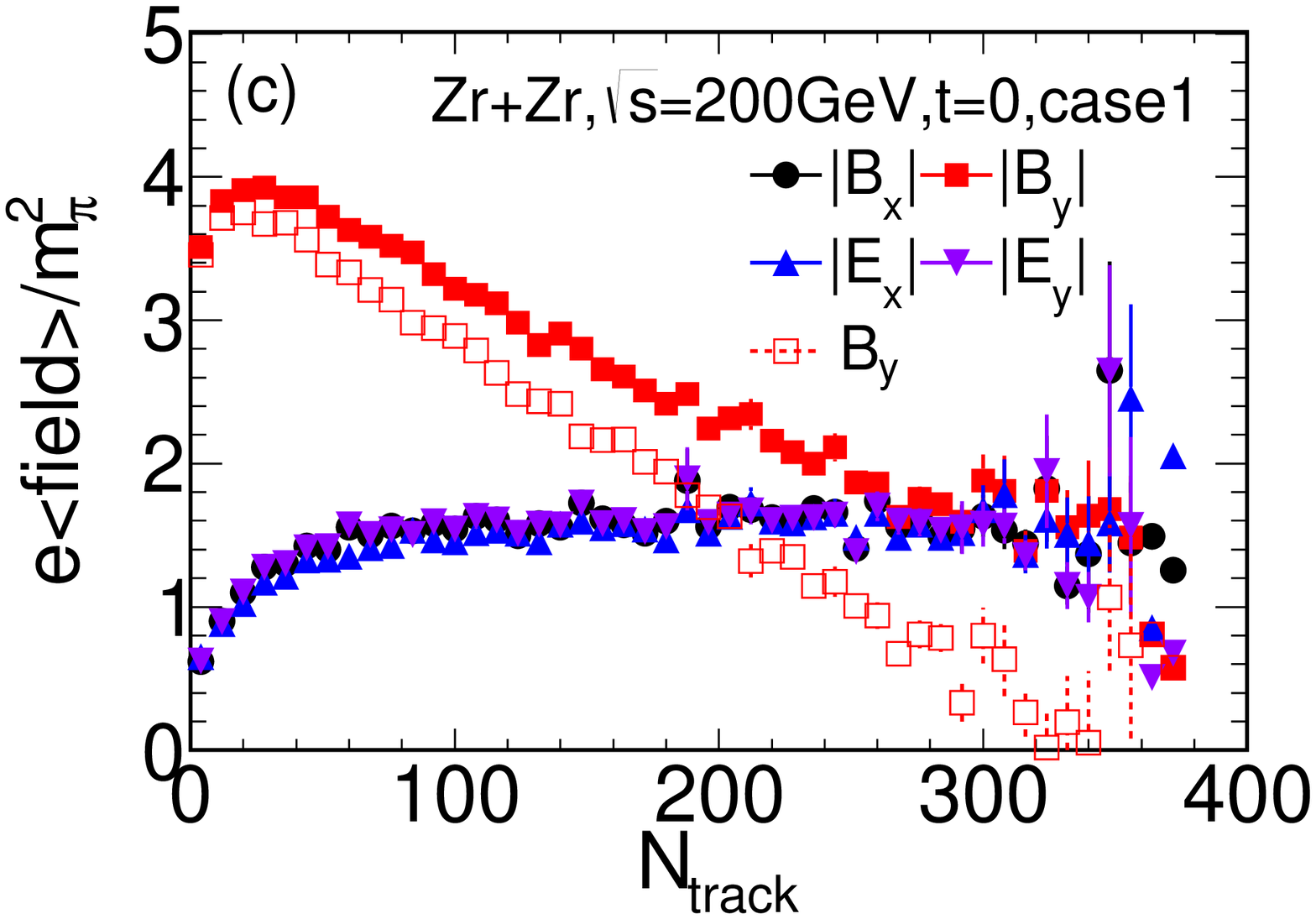}
    \label{fig:side:b}
  \end{minipage}
\caption{(Color online) The electromagnetic fields at $t = 0$ and ${\bf r}={\bf 0}$ as functions of (a) impact parameter $b$, (b) $N_{\rm part}$ and, (c) $N_{\rm track}$ in $_{40}^{96}\textrm{Zr}+\,_{40}^{96}\textrm{Zr}$ collisions at $\sqrt s = 200$ GeV for case 1.}
\label{fig:Zr1EB}
\end{figure*}

\begin{figure*}[htb]
  \begin{minipage}[t]{0.333\linewidth}
    \includegraphics[width=0.95\textwidth]{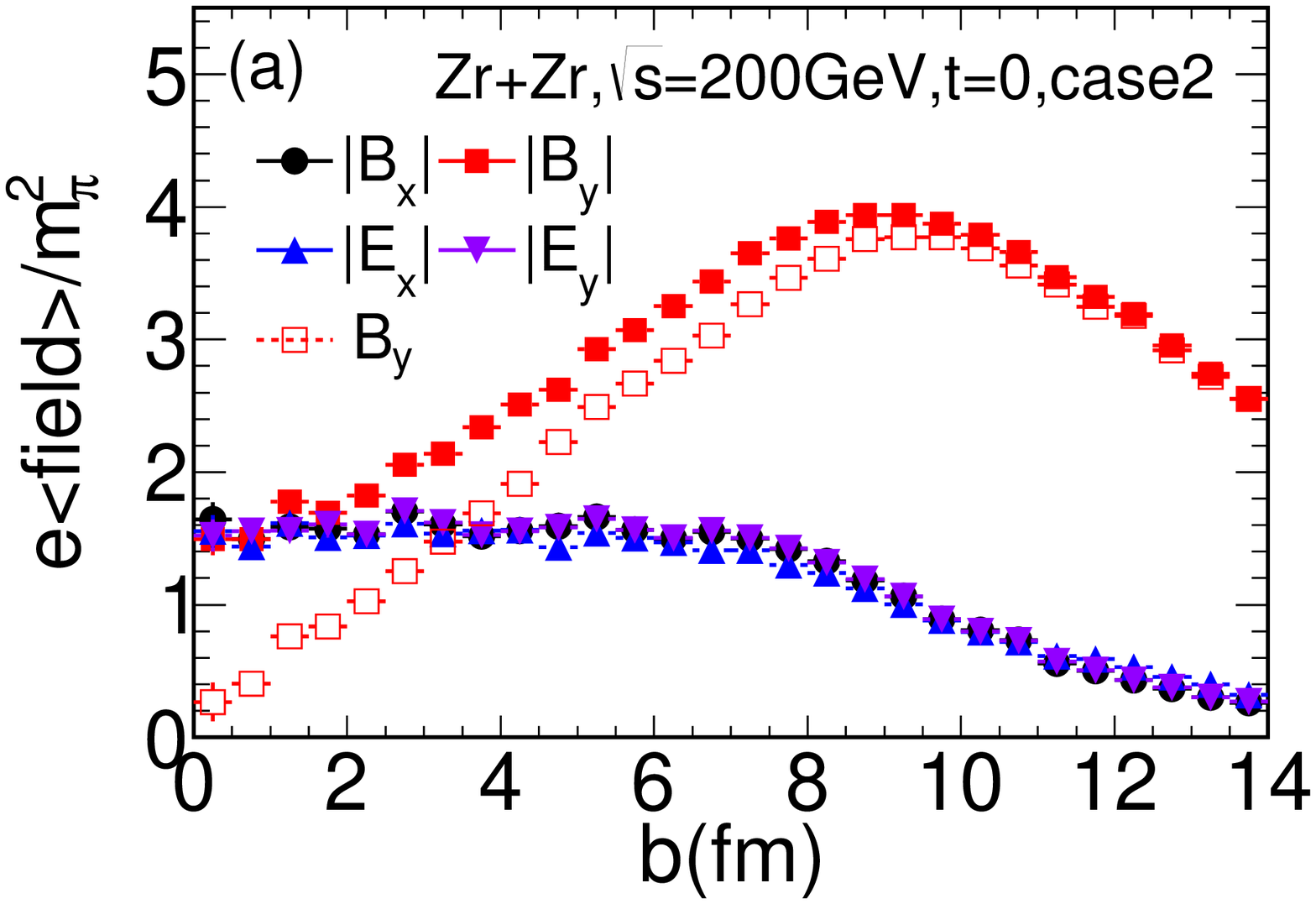}
    \label{fig:side:a}
  \end{minipage}%
  \begin{minipage}[t]{0.333\linewidth}
    \centering
    \includegraphics[width=0.95\textwidth]{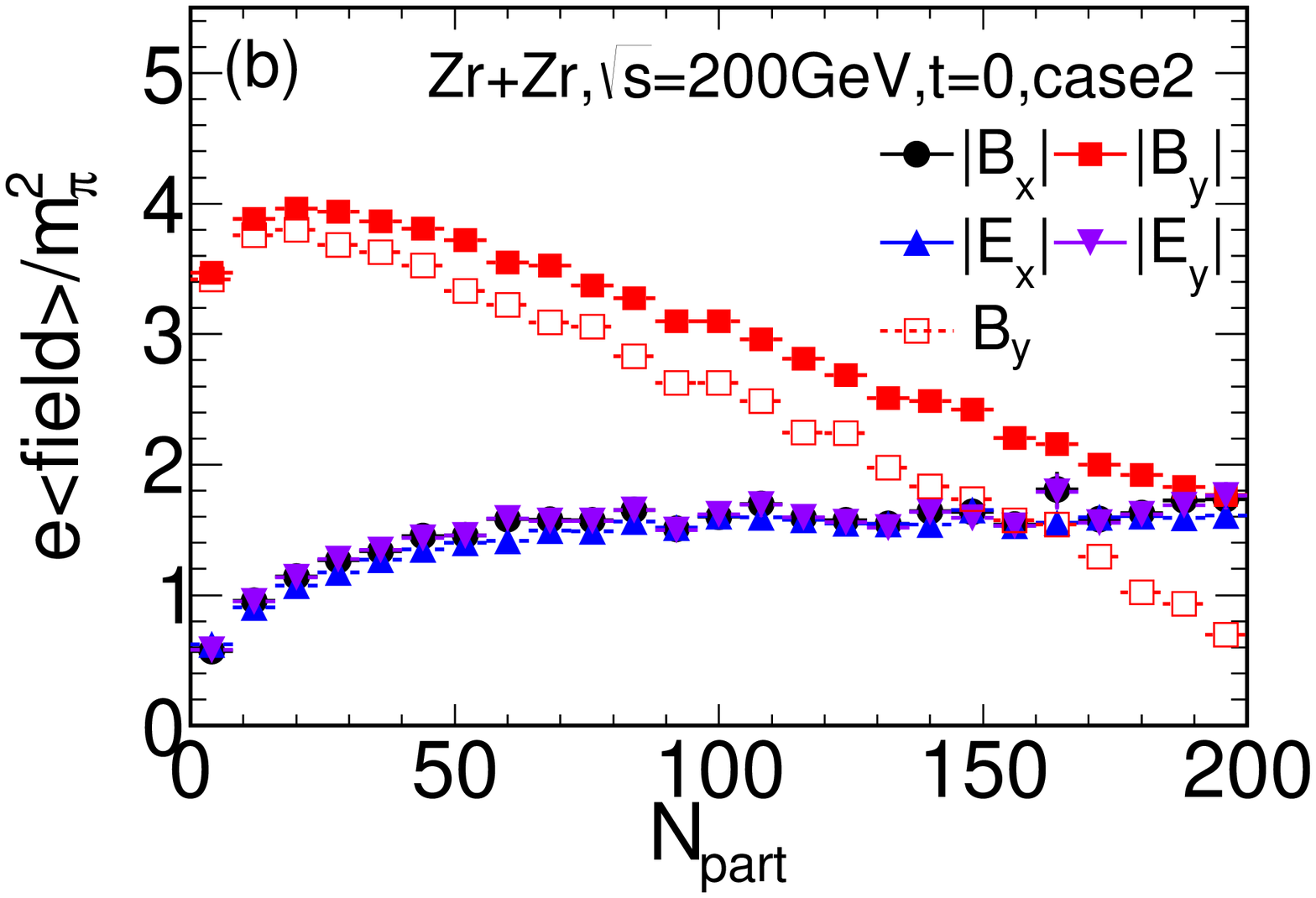}
    \label{fig:side:b}
  \end{minipage}%
  \begin{minipage}[t]{0.333\linewidth}
    \centering
    \includegraphics[width=0.95\textwidth]{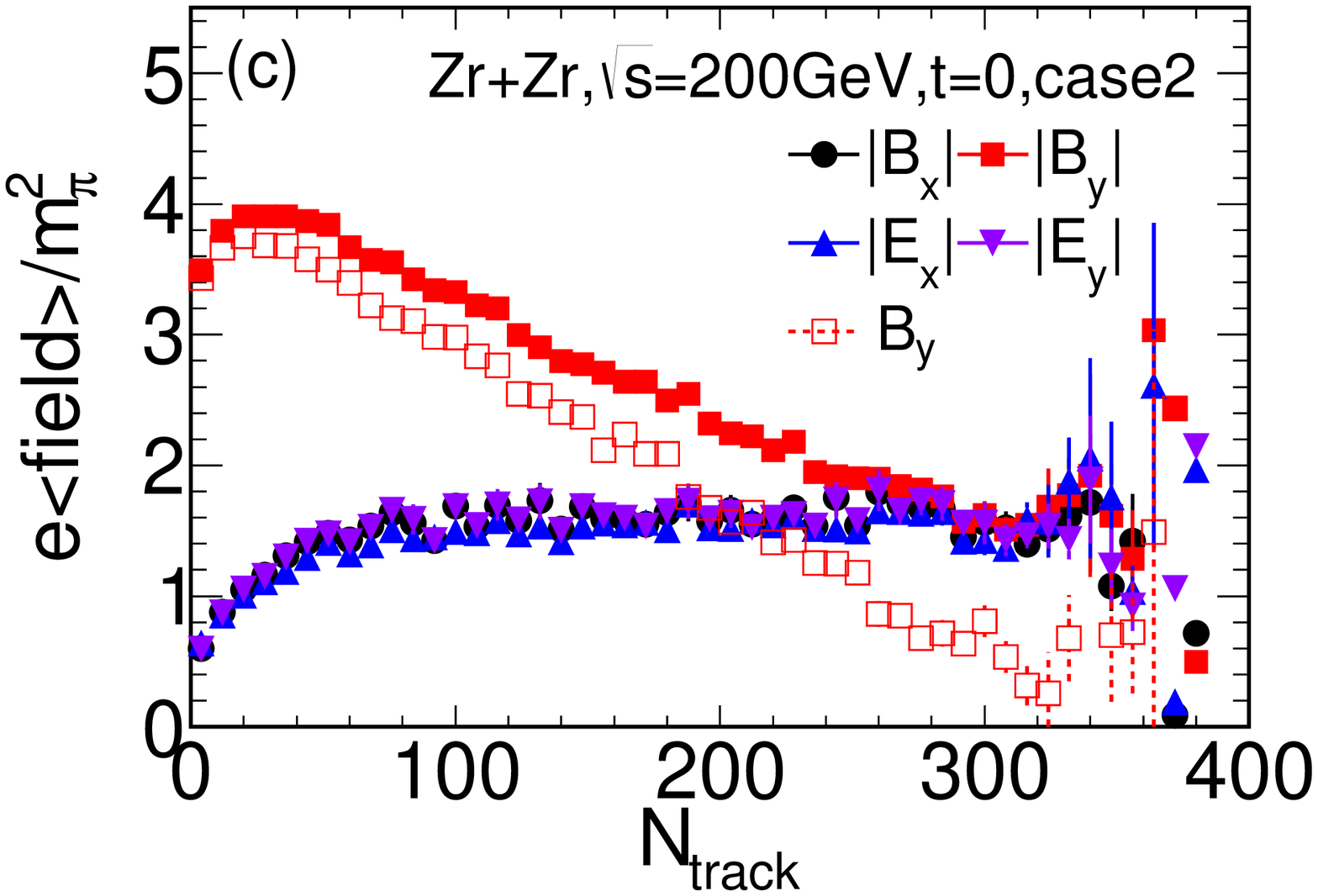}
    \label{fig:side:b}
  \end{minipage}
\caption{(Color online) The electromagnetic fields at $t = 0$ and ${\bf r}={\bf 0}$ as functions of (a) impact parameter $b$, (b) $N_{\rm part}$ and, (c) $N_{\rm track}$ in $_{40}^{96}\textrm{Zr}+\,_{40}^{96}\textrm{Zr}$ collisions at $\sqrt s = 200$ GeV for case 2.}
\label{fig:Zr2EB}
\end{figure*}

Figure~\ref{fig:Ru1EB} shows the electromagnetic fields in $_{44}^{96}\textrm{Ru}+\,_{44}^{96}\textrm{Ru}$ collisions at ${\bf r}$ = ${\bf 0}$ and $t$ = 0 at $\sqrt{s}$ = 200 GeV for case 1 where the panels $\rm(a)-(c)$ show the impact parameter $b$, $N_{\rm part}$ and $N_{\rm track}$ dependencies of electromagnetic fields, respectively. For the number of charged particles, $N_{\rm track}$, we set $|\eta| < 0.5$ and $p_{T} > 0.15$ GeV/c at the RHIC energy to match the STAR Collaboration acceptance. A point worth emphasizing is that $b$ and $N_{\rm part}$ are usually used in the model, whereas centrality and $N_{\rm track}$ are often used in experiments. We can easily find that the magnetic fields are almost zero in most central events and have the maximum at some peripheral events, which indicate we should search for the CME signals in peripheral collisions. Meanwhile, the average of the absolute value of electric fields gradually decreases as centrality increases. These results are similar to the results of electromagnetic fields for Au+Au collisions in shape, see Ref.~\cite{Zhao:2017rpf}. Because the radius of the Ru nucleus is smaller than that of the Au nucleus, the maximum of the magnetic fields is found in about $b$ = 9 fm for $_{44}^{96}\textrm{Ru}+\,_{44}^{96}\textrm{Ru}$ collisions, and it is about $b$ = 12 fm for Au+Au collisions. What is more, the Ru nucleus has less protons than the Au nucleus, so the magnitudes of electromagnetic fields for Ru+Ru collisions are smaller than those for Au+Au collisions. Figure~\ref{fig:Ru2EB} shows the electromagnetic fields in $_{44}^{96}\textrm{Ru}+\,_{44}^{96}\textrm{Ru}$ collisions at ${\bf r}$ = ${\bf 0}$ and $t$ = 0 at $\sqrt{s}$ = 200 GeV for case 2. We can see that the electromagnetic fields of case 2 look almost identical with case 1. Similarly, Figs.~\ref{fig:Zr1EB} and~\ref{fig:Zr2EB} show the results of electromagnetic fields in $_{40}^{96}\textrm{Zr}+\,_{40}^{96}\textrm{Zr}$ collisions for case 1 and case 2, respectively. (Note that the $\langle|B_{x}|\rangle$ and $\langle|E_{x, y}|\rangle$ look overlapped in Figs.~\ref{fig:Ru1EB} $-$~\ref{fig:Zr2EB}.)

\begin{figure*}[htb]
  \setlength{\abovecaptionskip}{0pt}
  \setlength{\belowcaptionskip}{8pt}\centerline{\includegraphics[scale=0.9]{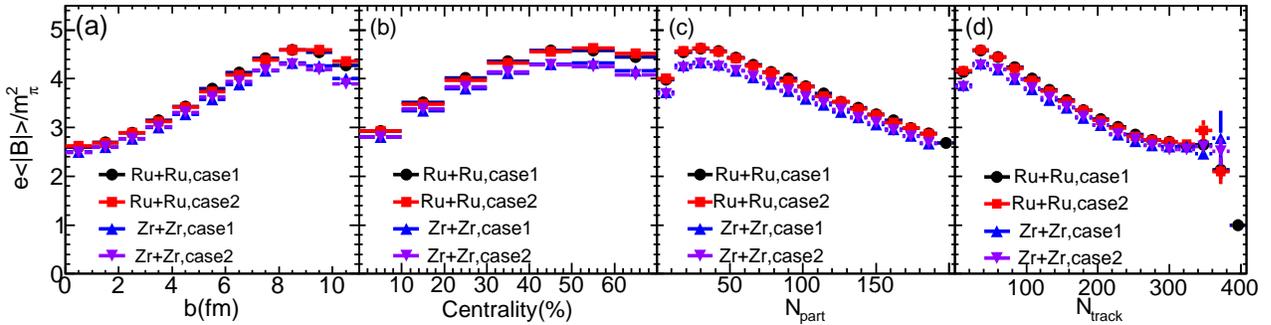}}
\caption{(Color online) The magnetic field at $t = 0$ and ${\bf r}={\bf 0}$ as functions of (a) impact parameter $b$, (b) centrality, (c) $N_{\rm part}$, and (d) $N_{\rm track}$ in $_{44}^{96}\textrm{Ru}+\,_{44}^{96}\textrm{Ru}$ collisions and $_{40}^{96}\textrm{Zr}+\,_{40}^{96}\textrm{Zr}$ collisions for case 1 and case 2 at $\sqrt s = 200$ GeV.}\label{fig:BB}
\end{figure*}

\begin{figure*}[htb]
  \setlength{\abovecaptionskip}{0pt}
  \setlength{\belowcaptionskip}{8pt}\centerline{\includegraphics[scale=0.9]{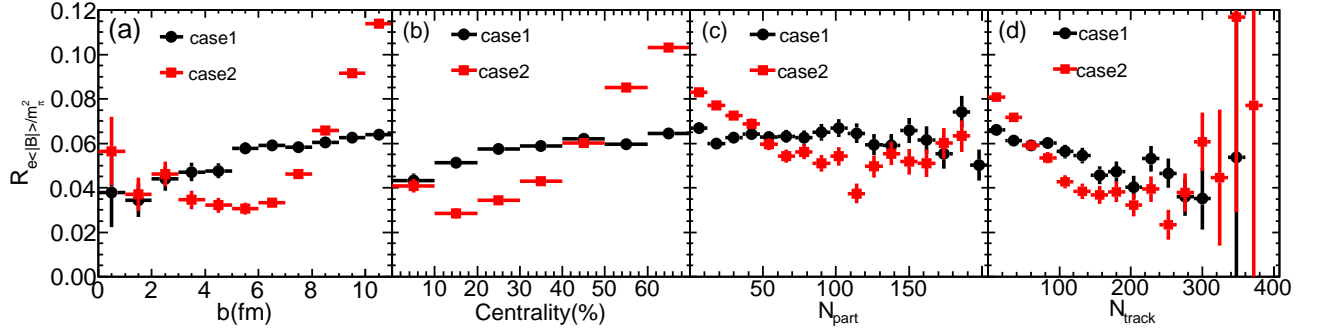}}
\caption{(Color online) The relative ratio of magnetic field as functions of (a) impact parameter $b$, (b) centrality, (c) $N_{\rm part}$, and (d) $N_{\rm track}$ in isobaric collisions at $\sqrt s = 200$ GeV for case 1 and case 2.}\label{fig:RBB}
\end{figure*}

Figure~\ref{fig:BB} shows the absolute value of the magnetic field in $_{44}^{96}\textrm{Ru}+\,_{44}^{96}\textrm{Ru}$ collisions for case 1 and case 2 and in $_{40}^{96}\textrm{Zr}+\,_{40}^{96}\textrm{Zr}$ collisions for case 1 and case 2 as functions of $b$, centrality, $N_{\rm part}$ and $N_{\rm track}$. We can clearly see that the magnetic field in $_{44}^{96}\textrm{Ru}+\,_{44}^{96}\textrm{Ru}$ collisions is larger than that in $_{40}^{96}\textrm{Zr}+\,_{40}^{96}\textrm{Zr}$ collisions, and for both case 1 and case 2. The magnitude of the magnetic field is almost the same between two cases for given isobaric collisions. In order to find the discrepancy between $_{44}^{96}\textrm{Ru}+\,_{44}^{96}\textrm{Ru}$ collisions and $_{40}^{96}\textrm{Zr}+\,_{40}^{96}\textrm{Zr}$ collisions, we plot the relative ratio between the two collisions as defined by Eq.~(\ref{ratio}). In Fig.~\ref{fig:RBB}, we can find the relative difference in $\bf B$ between $_{44}^{96}\textrm{Ru}+\,_{44}^{96}\textrm{Ru}$ collisions and $_{40}^{96}\textrm{Zr}+\,_{40}^{96}\textrm{Zr}$ collisions for case 1 is about 4$\%$ in central events and increases to 6$\%$ in peripheral events. However, the relative difference for case 2 is about 4$\%$ in central events and gently decreases to 3$\%$ in mid-central events then increases to 11.5$\%$ in peripheral events. Note that our relative difference of $\bf B$ for case 1 is similar to Refs.~\cite{Huang:2017azw,Deng:2018dut}. It is easy to be understood that the electromagnetic fields of $_{40}^{96}\textrm{Zr}+\,_{40}^{96}\textrm{Zr}$ collisions are smaller than $_{44}^{96}\textrm{Ru}+\,_{44}^{96}\textrm{Ru}$ collisions because they have less protons. Unquestionably, the difference of magnetic fields is vital for measuring the CME and we indeed find differences in the magnetic fields between two isobaric collisions. Furthermore, we measure the CME signal as mentioned above by using the correlator $\Delta \gamma$. Because $\Delta \gamma\propto\langle(eB/m_{\pi}^{2})^{2}{\rm cos}\ 2(\Psi_B - \Psi_{2})\rangle$ has similar flow due to the same atomic number, therefore it is key to check how different the $\langle{\rm cos}\ 2(\Psi_B - \Psi_{2})\rangle$ are between two isobaric collisions, which will be discussed next.

\subsection{Correlation between magnetic field and participant plane $\Psi_{2}$ in isobaric collisions}
\label{resultsC}

\begin{figure*}[htb]
  \setlength{\abovecaptionskip}{0pt}
  \setlength{\belowcaptionskip}{8pt}\centerline{\includegraphics[scale=0.6]{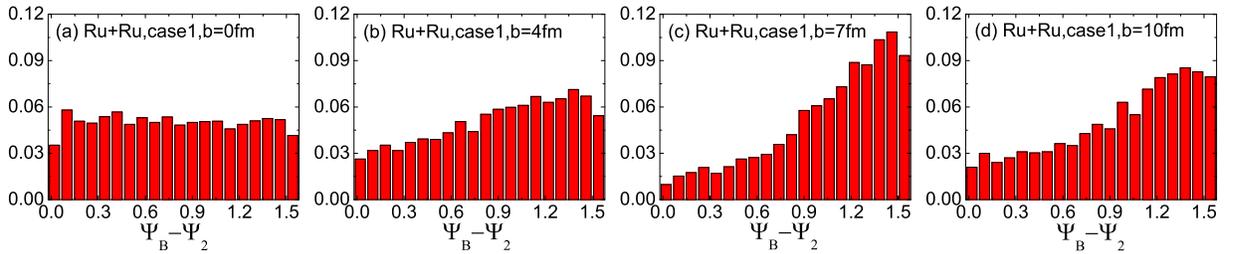}}
\caption{(Color online) The event-by-event histograms of $\Psi_{B}-\Psi_2$ at impact parameters $b =$ 0, 4, 7, and 10 fm in $_{44}^{96}\textrm{Ru}+\,_{44}^{96}\textrm{Ru}$ collisions at $\sqrt s = 200$ GeV for case 1. Here, $\Psi_{B}$ is the azimuthal direction of the {\bf B} field (at $t = 0$ and ${\bf r}={\bf 0}$), and $\Psi_2$ is the second harmonic participant plane.}\label{fig:Rupsidiff}
\end{figure*}

\begin{figure*}[htb]
  \setlength{\abovecaptionskip}{0pt}
  \setlength{\belowcaptionskip}{8pt}\centerline{\includegraphics[scale=0.6]{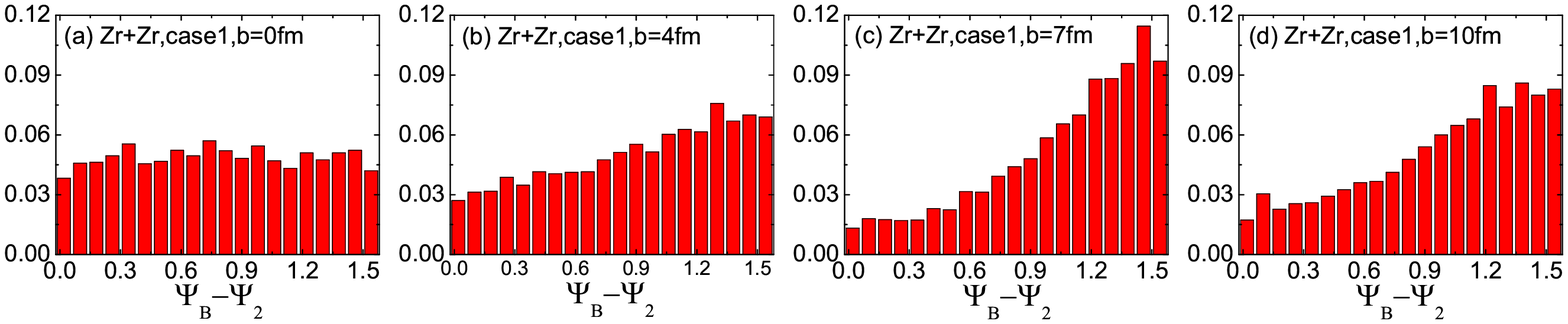}}
\caption{(Color online) The event-by-event histograms of $\Psi_{B}-\Psi_2$ at impact parameters $b =$ 0, 4, 7, and 10 fm in $_{40}^{96}\textrm{Zr}+\,_{40}^{96}\textrm{Zr}$ collisions at $\sqrt s = 200$ GeV for case 1. Here, $\Psi_{B}$ is the azimuthal direction of the {\bf B} field (at $t = 0$ and ${\bf r}={\bf 0}$), and $\Psi_2$ is the second-harmonic participant plane.}\label{fig:Zrpsidiff}
\end{figure*}

\begin{figure*}[htb]
  \setlength{\abovecaptionskip}{0pt}
  \setlength{\belowcaptionskip}{8pt}\centerline{\includegraphics[scale=0.6]{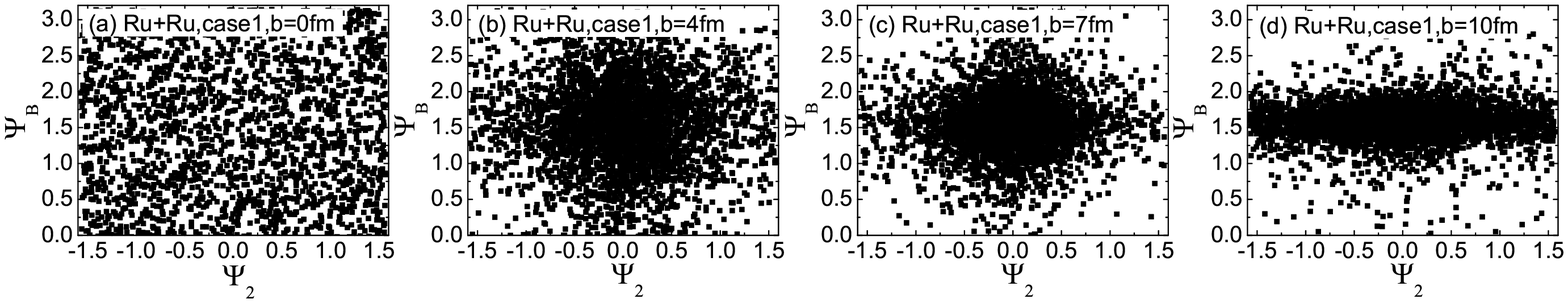}}
\caption{The scatter plots on $\Psi_{B}-\Psi_2$ plane at impact parameters $b =$ 0, 4, 7, and 10 fm in $_{44}^{96}\textrm{Ru}+\,_{44}^{96}\textrm{Ru}$ collisions at $\sqrt s = 200$ GeV for case 1, where $\Psi_{B}$ is the azimuthal direction of the {\bf B} field (at $t = 0$ and ${\bf r}={\bf 0}$), and $\Psi_2$ is the second-harmonic participant plane.}\label{fig:Ru2Dplot}
\end{figure*}

\begin{figure*}[htb]
  \setlength{\abovecaptionskip}{0pt}
  \setlength{\belowcaptionskip}{8pt}\centerline{\includegraphics[scale=0.6]{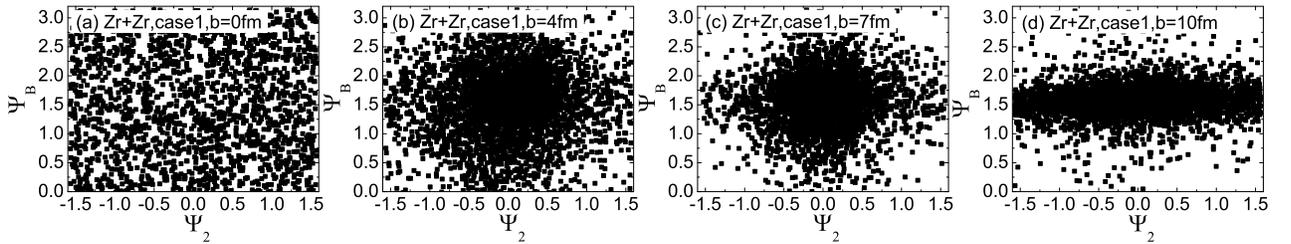}}
\caption{The scatter plots on $\Psi_{B}-\Psi_2$ plane at impact parameters $b =$ 0, 4, 7, and 10 fm in $_{40}^{96}\textrm{Zr}+\,_{40}^{96}\textrm{Zr}$ collisions at $\sqrt s = 200$ GeV for case 1, where $\Psi_{B}$ is the azimuthal direction of the {\bf B} field (at $t = 0$ and ${\bf r}={\bf 0}$), and $\Psi_2$ is the second-harmonic participant plane.}\label{fig:Zr2Dplot}
\end{figure*}

As the chiral anomalous  effects always occur either along or perpendicular to the magnetic-field direction, it is important to find an experimental way to determine the direction of the magnetic field. With the help of finite correlation between $\Psi_B $ and $\Psi_{2}$, ones fortunately are capable of accessing the magnetic field direction and then measuring the CME. In Figs.~\ref{fig:Rupsidiff} and~\ref{fig:Zrpsidiff} we plot the accumulated histograms of $\Psi_B - \Psi_{2}$ at $b$ = 0, 4, 7, and 10 fm in $_{44}^{96}\textrm{Ru}+\,_{44}^{96}\textrm{Ru}$ collisions for case 1 and in $_{40}^{96}\textrm{Zr}+\,_{40}^{96}\textrm{Zr}$ collisions for case 1, respectively. For $b$ = 0 fm, the histograms of $\Psi_B - \Psi_{2}$ are basically flat indicating that $\Psi_B $ and $\Psi_{2}$ are uncorrelated. For $b$ = 4, 7, and 10 fm, the histogram has a shape peaking at $\Psi_B - \Psi_{2} = \pi/2$ with corresponding widths. This implies some correlation exists between $\Psi_{B}$ and $\Psi_{2}$. Figures~\ref{fig:Ru2Dplot} and~\ref{fig:Zr2Dplot} show the corresponding two-dimensional correlation distributions for $\Psi_B $ and $\Psi_{2}$ in $_{44}^{96}\textrm{Ru}+\,_{44}^{96}\textrm{Ru}$ collisions and in $_{40}^{96}\textrm{Zr}+\,_{40}^{96}\textrm{Zr}$ collisions for case 1, respectively. For $b$ = 0 fm, the events are almost uniformly distributed, indicating a negligible correlation between $\Psi_B $ and $\Psi_{2}$. For $b$= 4, 7, and 10 fm, the event distributions evidently concentrate around $\Psi_B - \Psi_{2}$ = $\pi/2$, indicating an existing correlation between the two angles.

\begin{figure*}[htb]
  \setlength{\abovecaptionskip}{0pt}
  \setlength{\belowcaptionskip}{8pt}\centerline{\includegraphics[scale=0.9]{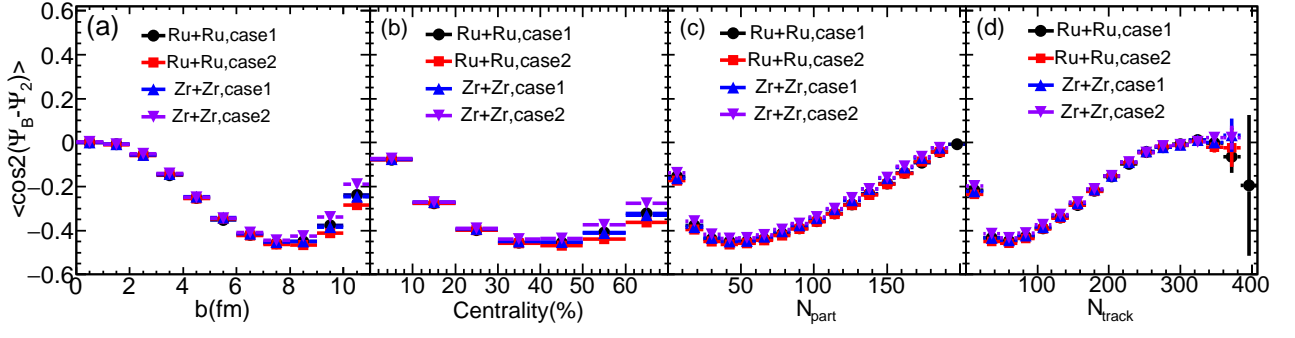}}
\caption{(Color online) The correlation $\langle{\rm cos}\ 2(\Psi_B - \Psi_{2})\rangle$ as functions of (a) impact parameter $b$, (b) centrality, (c) $N_{\rm part}$, and (d) $N_{\rm track}$ in $_{44}^{96}\textrm{Ru}+\,_{44}^{96}\textrm{Ru}$ collisions and $_{40}^{96}\textrm{Zr}+\,_{40}^{96}\textrm{Zr}$ collisions at $\sqrt s = 200$ GeV for case 1 and case 2.}\label{fig:cos}
\end{figure*}
\begin{figure*}[htb]
  \setlength{\abovecaptionskip}{0pt}
  \setlength{\belowcaptionskip}{8pt}\centerline{\includegraphics[scale=0.9]{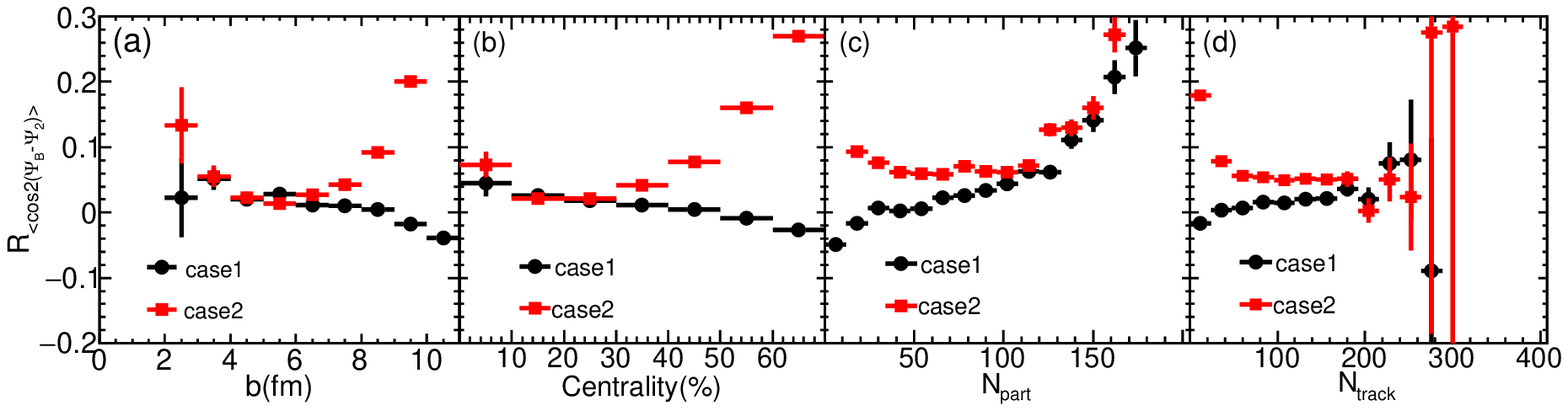}}
\caption{(Color online) The relative ratio of $\langle{\rm cos}\ 2(\Psi_B - \Psi_{2})\rangle$ as functions of (a) impact parameter $b$, (b) centrality, (c) $N_{\rm part}$, and (d) $N_{\rm track}$ in isobaric collisions at $\sqrt s = 200$ GeV for case 1 and case 2.}\label{fig:Rcos}
\end{figure*}

\begin{figure*}[htb]
  \setlength{\abovecaptionskip}{0pt}
  \setlength{\belowcaptionskip}{8pt}\centerline{\includegraphics[scale=0.9]{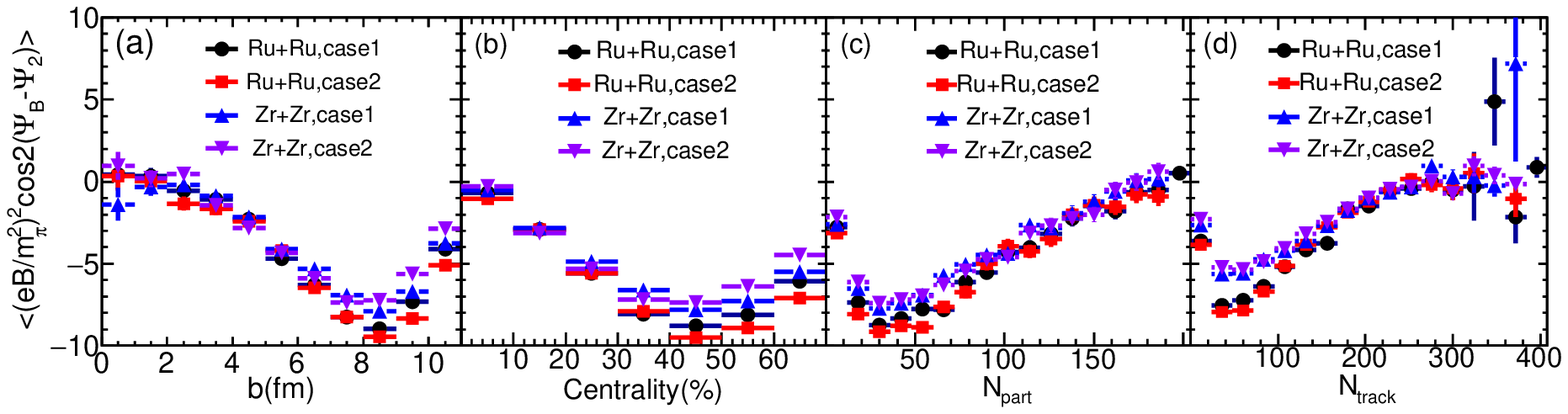}}
\caption{(Color online) The correlation $\langle(eB/m_{\pi}^{2})^{2}{\rm cos}\ 2(\Psi_B - \Psi_{2})\rangle$ as functions of (a) impact parameter $b$, (b) centrality, (c) $N_{\rm part}$, and (d) $N_{\rm track}$ in $_{44}^{96}\textrm{Ru}+\,_{44}^{96}\textrm{Ru}$ collisions and $_{40}^{96}\textrm{Zr}+\,_{40}^{96}\textrm{Zr}$ collisions at $\sqrt s = 200$ GeV for case 1 and case 2.}\label{fig:BBcos}
\end{figure*}
\begin{figure*}[htb]
  \setlength{\abovecaptionskip}{0pt}
  \setlength{\belowcaptionskip}{8pt}\centerline{\includegraphics[scale=0.9]{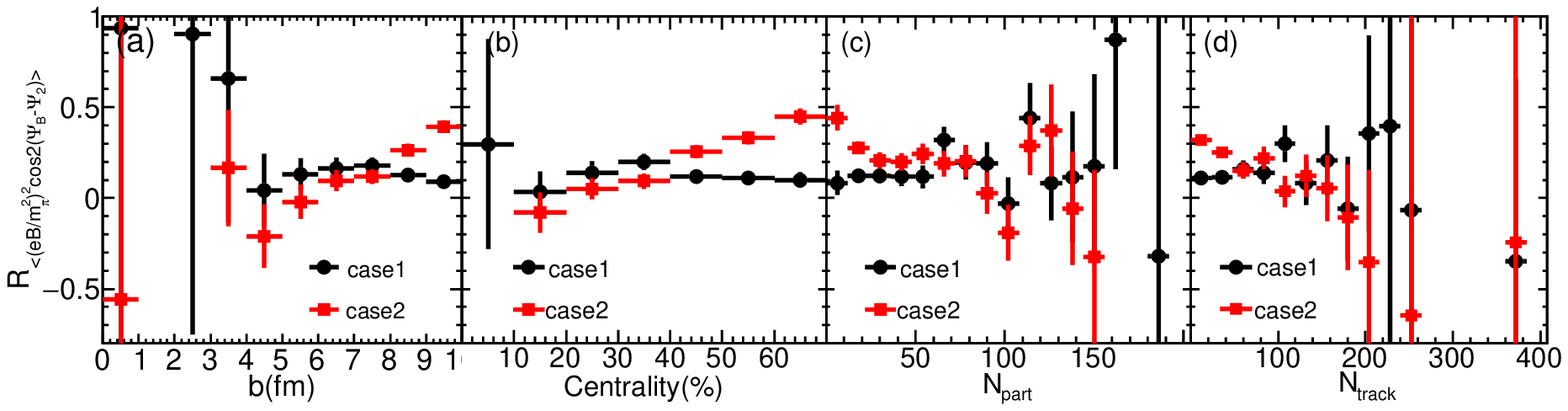}}
\caption{(Color online) The relative ratio of $\langle(eB/m_{\pi}^{2})^{2}{\rm cos}\ 2(\Psi_B - \Psi_{2})\rangle$ as functions of (a) impact parameter $b$, (b) centrality, (c) $N_{\rm part}$, and (d) $N_{\rm track}$ in isobaric collisions at $\sqrt s = 200$ GeV for case 1 and case 2.}\label{fig:RBBcos}
\end{figure*}

Figure~\ref{fig:cos} shows that the correlation between magnetic field and participant plane $\Psi_{2}$ in $_{44}^{96}\textrm{Ru}+\,_{44}^{96}\textrm{Ru}$ collisions and in $_{40}^{96}\textrm{Zr}+\,_{40}^{96}\textrm{Zr}$ collisions for case 1 and case 2 as functions of $b$, centrality, $N_{\rm part}$ and $N_{\rm track}$. Obviously, the correlation of $\langle{\rm cos}\ 2(\Psi_B - \Psi_{2})\rangle$ depended on centrality and these results are consistent with Figs.~\ref{fig:Rupsidiff} -~\ref{fig:Zr2Dplot}. In most central events and most peripheral events, $\langle{\rm cos}\ 2(\Psi_B - \Psi_{2})\rangle$ is almost zero, due to large fluctuations. However, $\langle{\rm cos}\ 2(\Psi_B - \Psi_{2})\rangle$ has a maximum about $-0.5$ for both $_{44}^{96}\textrm{Ru}+\,_{44}^{96}\textrm{Ru}$ collisions and $_{40}^{96}\textrm{Zr}+\,_{40}^{96}\textrm{Zr}$ collisions at $b = 6-9$ fm. The correlations for the four cases look quite similar. Then, we also take the relative ratios between $_{44}^{96}\textrm{Ru}+\,_{44}^{96}\textrm{Ru}$ collisions and $_{40}^{96}\textrm{Zr}+\,_{40}^{96}\textrm{Zr}$ collisions for case 1 and case 2, which are shown in Fig.~\ref{fig:Rcos}. We can see for case 1, the relative ratio of $\langle{\rm cos}\ 2(\Psi_B - \Psi_{2})\rangle$ between $_{44}^{96}\textrm{Ru}+\,_{44}^{96}\textrm{Ru}$ collisions and $_{40}^{96}\textrm{Zr}+\,_{40}^{96}\textrm{Zr}$ collisions is about 5$\%$ in most central bins then decreases to $-2\%$ in most peripheral bins. For case 2, the relative ratio of $\langle{\rm cos}\ 2(\Psi_B - \Psi_{2})\rangle$ is concave and about 7$\%$ in most central bins and then increases to about 27$\%$ in most peripheral bins. In peripheral bins, one can see that the relative differences of $\langle{\rm cos}\ 2(\Psi_B - \Psi_{2})\rangle$ for case 1 and case 2 differ a lot, which is actually caused by the deformation, i.e., the larger deformation and the weaker correlation $\langle{\rm cos}\ 2(\Psi_B - \Psi_{2})\rangle$ for case 2 as shown in Fig.~\ref{fig:cos}.

Figure~\ref{fig:BBcos} shows that the correlation $\langle(eB/m_{\pi}^{2})^{2}{\rm cos}\ 2(\Psi_B - \Psi_{2})\rangle$ in $_{44}^{96}\textrm{Ru}+\,_{44}^{96}\textrm{Ru}$ collisions and in $_{40}^{96}\textrm{Zr}+\,_{40}^{96}\textrm{Zr}$ collisions for case 1 and case 2 as functions of $b$, centrality, $N_{\rm part}$ and $N_{\rm track}$. It shows a distinct difference of $\langle(eB/m_{\pi}^{2})^{2}{\rm cos}\ 2(\Psi_B - \Psi_{2})\rangle$ between $_{44}^{96}\textrm{Ru}+\,_{44}^{96}\textrm{Ru}$ collisions and $_{40}^{96}\textrm{Zr}+\,_{40}^{96}\textrm{Zr}$ collisions. By contrasting with Figs.~\ref{fig:BB} and~\ref{fig:cos}, we can see that it is caused by both the magnetic field and the correlation. Following the same way, the relative ratios of $\langle(eB/m_{\pi}^{2})^{2}{\rm cos}\ 2(\Psi_B - \Psi_{2})\rangle$ between $_{44}^{96}\textrm{Ru}+\,_{44}^{96}\textrm{Ru}$ collisions and $_{40}^{96}\textrm{Zr}+\,_{40}^{96}\textrm{Zr}$ collisions are presented in Fig.~\ref{fig:RBBcos}. From Fig.~\ref{fig:RBBcos}(b), we can clearly see for case 1, the relative ratio is flat near 10$\%$. But for case 2, the relative ratio shows a clear increasing trend from central to peripheral events. By comparing the results from Figs.~\ref{fig:RBB},~\ref{fig:Rcos}, and~\ref{fig:RBBcos}, we find the relative ratio of $\langle(eB/m_{\pi}^{2})^{2} {\rm cos}\ 2(\Psi_B - \Psi_{2})\rangle$ is larger than the relative ratio of $\langle{\rm cos}\ 2(\Psi_B - \Psi_{2})\rangle$, which indicates that the magnetic field plays an important role on the CME observable. All the relative ratios between $_{44}^{96}\textrm{Ru}+\,_{44}^{96}\textrm{Ru}$ collisions and $_{40}^{96}\textrm{Zr}+\,_{40}^{96}\textrm{Zr}$ collisions for case 1 are similar to case 2 for mid central events, but in peripheral events the relative ratios between $_{44}^{96}\textrm{Ru}+\,_{44}^{96}\textrm{Ru}$ collisions and $_{40}^{96}\textrm{Zr}+\,_{40}^{96}\textrm{Zr}$ collisions for case 1 are less than that for case 2. Our results indicate that the deformation has almost no effect on $\langle{\rm cos}\ 2(\Psi_B - \Psi_{2})\rangle$ and $\langle (eB/m_{\pi}^{2})^{2}{\rm cos}\ 2(\Psi_B - \Psi_{2})\rangle$ in central and mid central events, but can not be neglected in peripheral events.

\subsection{Correlation between magnetic field and spectator plane $\Psi_{2}^{\rm SP}$ in isobaric collisions}
\label{resultsD}
The previous subsection shows the results from the correlation between magnetic field and participant plane $\Psi_{2}$, now we show the results from the correlation  between magnetic field and spectator plane $\Psi_{2}^{\rm SP}$.
\begin{figure*}[htb]
  \setlength{\abovecaptionskip}{0pt}
  \setlength{\belowcaptionskip}{8pt}\centerline{\includegraphics[scale=0.9]{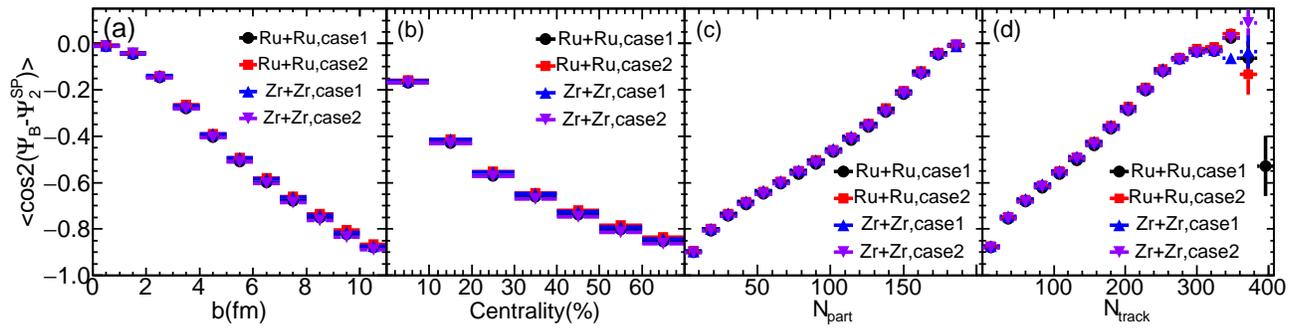}}
\caption{(Color online) The correlation $\langle{\rm cos}\ 2(\Psi_B - \Psi_{2}^{\rm SP})\rangle$ as functions of (a) impact parameter $b$, (b) centrality, (c) $N_{\rm part}$, and (d) $N_{\rm track}$ in $_{44}^{96}\textrm{Ru}+\,_{44}^{96}\textrm{Ru}$ collisions and $_{40}^{96}\textrm{Zr}+\,_{40}^{96}\textrm{Zr}$ collisions at $\sqrt s = 200$ GeV for case 1 and case 2.}\label{fig:spcos}
\end{figure*}

Figure~\ref{fig:spcos} shows the correlations between magnetic field direction $\Psi_B$ and spectator plane $\Psi_{2}^{\rm SP}$, $\langle{\rm cos}\ 2(\Psi_B - \Psi_{2}^{\rm SP})\rangle$, in $_{44}^{96}\textrm{Ru}+\,_{44}^{96}\textrm{Ru}$ collisions and  $_{40}^{96}\textrm{Zr}+\,_{40}^{96}\textrm{Zr}$ collisions for case 1 and case 2 as functions of $b$, centrality, $N_{\rm part}$, and $N_{\rm track}$. Compared to Fig.~\ref{fig:cos}, the correlation of $\langle{\rm cos}\ 2(\Psi_B - \Psi_{2}^{\rm SP})\rangle$ is around two times larger than that between magnetic field and participant plane $\Psi_{2}$, $\langle{\rm cos}\ 2(\Psi_B - \Psi_{2})\rangle$. In peripheral collisions, this correlation is much stronger and approaching one.

\begin{figure*}[htb]
  \setlength{\abovecaptionskip}{0pt}
  \setlength{\belowcaptionskip}{8pt}\centerline{\includegraphics[scale=0.9]{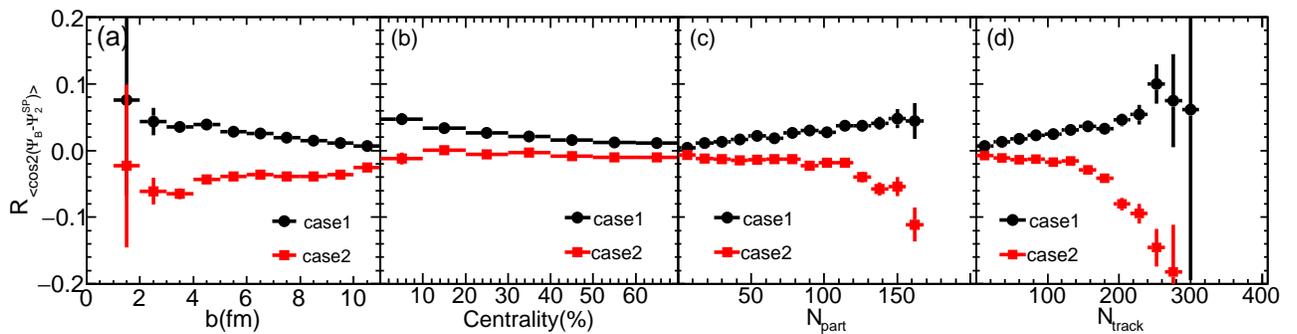}}
\caption{(Color online) The relative ratio of $\langle{\rm cos}\ 2(\Psi_B - \Psi_{2}^{\rm SP})\rangle$ as functions of (a) impact parameter $b$, (b) centrality, (c) $N_{\rm part}$, and (d) $N_{\rm track}$ in isobaric collisions at $\sqrt s = 200$ GeV for case 1 and case 2.}\label{fig:spRcos}
\end{figure*}

In the same way, we also take the relative ratio between $_{44}^{96}\textrm{Ru}+\,_{44}^{96}\textrm{Ru}$ collisions and $_{40}^{96}\textrm{Zr}+\,_{40}^{96}\textrm{Zr}$ collisions for case 1 and case 2, as shown in Fig.~\ref{fig:spRcos}. The relative ratios of $\langle{\rm cos}\ 2(\Psi_B - \Psi_{2}^{\rm SP})\rangle$ for case 1 are gradually decreased from around 5\% to 0. Compared with Fig.~\ref{fig:Rcos}, the relative ratios of $\langle{\rm cos}\ 2(\Psi_B - \Psi_{2}^{\rm SP})\rangle$ for both case 1 and case 2 are close to zero for noncentral collisions. This indicates that there is little difference in the terms of $\langle{\rm cos}\ 2(\Psi_B - \Psi_{2}^{\rm SP})\rangle$ between the two isobaric collisions for both cases for non-central collisions, thanks to the strong correlation between $\Psi_B$ and $\Psi_{2}^{\rm SP}$. It provides a natural advantage to detect the possible effects purely from the difference of magnetic fields, even with less influence of the deformation.

\begin{figure*}[htb]
  \setlength{\abovecaptionskip}{0pt}
  \setlength{\belowcaptionskip}{8pt}\centerline{\includegraphics[scale=0.9]{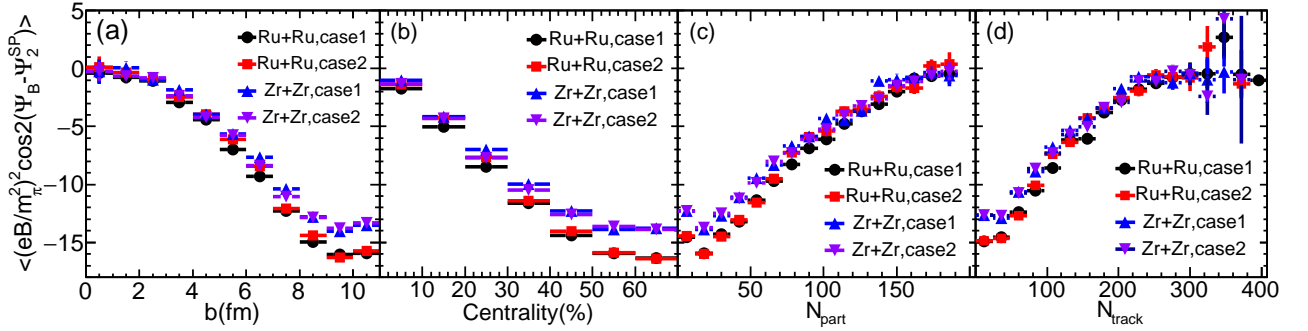}}
\caption{(Color online) The correlation $\langle{\rm cos}\ 2(\Psi_B - \Psi_{2}^{\rm SP})\rangle$ as functions of (a) impact parameter $b$, (b) centrality, (c) $N_{\rm part}$, and (d) $N_{\rm track}$ in $_{44}^{96}\textrm{Ru}+\,_{44}^{96}\textrm{Ru}$ collisions and $_{40}^{96}\textrm{Zr}+\,_{40}^{96}\textrm{Zr}$ collisions at $\sqrt s = 200$ GeV for case 1 and case 2.}\label{fig:spBBcos}
\end{figure*}
\begin{figure*}[htb]
  \setlength{\abovecaptionskip}{0pt}
  \setlength{\belowcaptionskip}{8pt}\centerline{\includegraphics[scale=0.9]{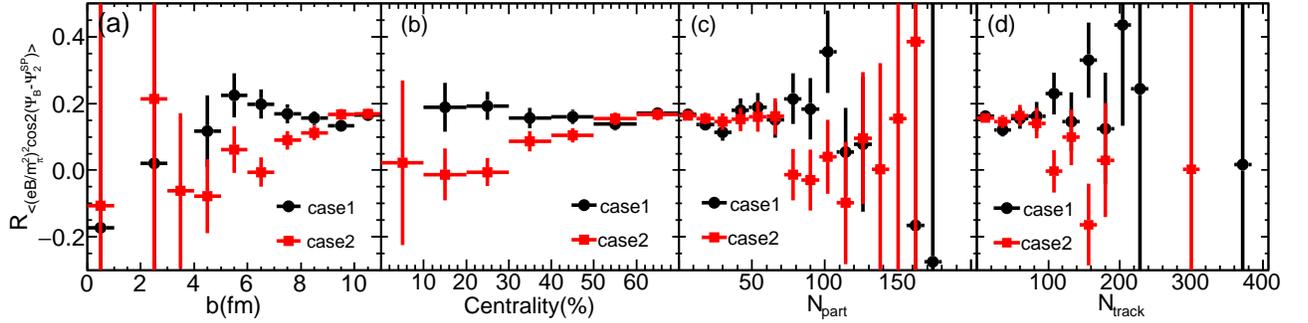}}
\caption{(Color online) The relative ratio of $\langle{\rm cos}\ 2(\Psi_B - \Psi_{2}^{\rm SP})\rangle$ as functions of (a) impact parameter $b$, (b) centrality, (c) $N_{\rm part}$, and (d) $N_{\rm track}$ in isobaric collisions at $\sqrt s = 200$ GeV for case 1 and case 2.}\label{fig:spRBBcos}
\end{figure*}

Figure~\ref{fig:spBBcos} shows that the correlation $\langle(eB/m_{\pi}^{2})^{2}{\rm cos}\ 2(\Psi_B - \Psi_{2}^{\rm SP})\rangle$ in $_{44}^{96}\textrm{Ru}+\,_{44}^{96}\textrm{Ru}$ collisions and $_{40}^{96}\textrm{Zr}+\,_{40}^{96}\textrm{Zr}$ collisions for case 1 and case 2 as functions of $b$, centrality, $N_{\rm part}$, and $N_{\rm track}$. Note that compared to Fig.~\ref{fig:BBcos}, the magnetic field is the same, but ${\rm cos}\ 2(\Psi_B - \Psi_{2}^{\rm SP})$ makes a difference. Because the magnetic field has a stronger correlation with the spectator plane than the participant plane, $\langle(eB/m_{\pi}^{2})^{2}{\rm cos}\ 2(\Psi_B - \Psi_{2}^{\rm SP})\rangle$ is stronger than $\langle(eB/m_{\pi}^{2})^{2}{\rm cos}\ 2(\Psi_B - \Psi_{2})\rangle$. The relative ratios of $\langle(eB/m_{\pi}^{2})^{2}{\rm cos}\ 2(\Psi_B - \Psi_{2}^{\rm SP})\rangle$ between $_{44}^{96}\textrm{Ru}+\,_{44}^{96}\textrm{Ru}$ collisions and $_{40}^{96}\textrm{Zr}+\,_{40}^{96}\textrm{Zr}$ collisions are presented in Fig.~\ref{fig:spRBBcos}. For case 1, the ratio fluctuates near 15$\%$ which is similar to the relative ratio of $\langle(eB/m_{\pi}^{2})^{2}{\rm cos}\ 2(\Psi_B - \Psi_{2})\rangle$. For case 2, the relative ratio increases from central to peripheral events which is similar to the trend of $\langle(eB/m_{\pi}^{2})^{2}{\rm cos}\ 2(\Psi_B - \Psi_{2})\rangle$, but the magnitude is reduced from 40\% to 20\% for the peripheral collisions.

\begin{figure*}[htb]
  \setlength{\abovecaptionskip}{0pt}
  \setlength{\belowcaptionskip}{8pt}\centerline{\includegraphics[scale=0.99]{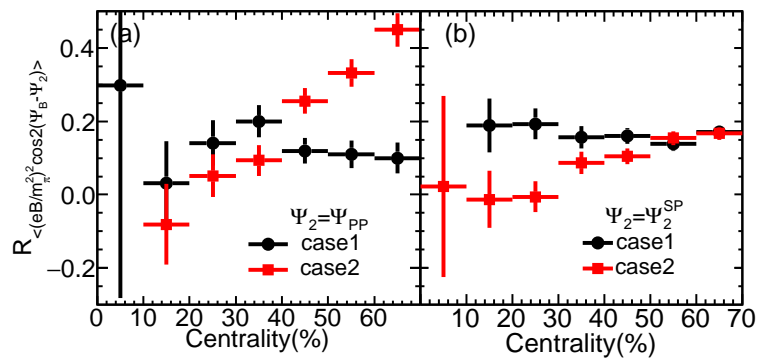}}
\caption {The relative ratios of (a) $\langle(eB/m_{\pi}^{2})^{2}{\rm cos}\ 2(\Psi_B - \Psi_{2})\rangle$ and (b) $\langle(eB/m_{\pi}^{2})^{2}{\rm cos}\ 2(\Psi_B - \Psi_{2}^{\rm SP})\rangle$ as functions of the centrality bin in isobaric collisions at $\sqrt s = 200$ GeV for case 1 and case 2.}
\label{fig:vscen}
\end{figure*}

Figure~\ref{fig:vscen} gives a direct comparison between $\langle(eB/m_{\pi}^{2})^{2}{\rm cos}\ 2(\Psi_B - \Psi_{2})\rangle$ and $\langle(eB/m_{\pi}^{2})^{2}{\rm cos}\ 2(\Psi_B - \Psi_{2}^{\rm SP})\rangle$ as functions of the centrality bin for case 1 and case 2. We find that the relative ratios of $\langle(eB/m_{\pi}^{2})^{2}{\rm cos}\ 2(\Psi_B - \Psi_{2})\rangle$ and $\langle(eB/m_{\pi}^{2})^{2}{\rm cos}\ 2(\Psi_B - \Psi_{2}^{\rm SP})\rangle$ for case 1 are similar, because the deformation difference is relatively weak for case 1. However, we observe that the two methods present different results for case 2, i.e., the relative ratio for the participant plane is larger than the relative ratio for the spectator plane. Based on the above results, we have already known that the correlation with the spectator plane is stronger than that with the participant plane. Therefore, $\langle(eB/m_{\pi}^{2})^{2}{\rm cos}\ 2(\Psi_B - \Psi_{2}^{\rm SP})\rangle$ is mainly affected by the magnetic field, however, $\langle(eB/m_{\pi}^{2})^{2}{\rm cos}\ 2(\Psi_B - \Psi_{2})\rangle$ is affected by both magnetic field and $\langle{\rm cos}\ 2(\Psi_B - \Psi_{2})\rangle$. It suggests that we can observe a much cleaner magnetic field effect of CME with the correlation $\Delta \gamma$ with respect to the spectator plane than that with respect to the participant plane.

\section{Conclusions}
\label{summary}
To summarize, we have utilized the AMPT model to investigate the properties of electromagnetic fields in isobaric $_{44}^{96}\textrm{Ru}+\,_{44}^{96}\textrm{Ru}$ collisions and $_{40}^{96}\textrm{Zr}+\,_{40}^{96}\textrm{Zr}$ collisions at the RHIC energy of $\sqrt{s}$=200 GeV. Meanwhile, the relative ratios of the magnetic fields are up to 10$\%$ for different centralities for case 1 and case 2.
Furthermore, the correlations $\langle{\rm cos}\ 2(\Psi_B - \Psi_{2}^{\rm SP})\rangle$ and $\langle(eB/m_{\pi}^{2})^{2}{\rm cos}\ 2(\Psi_B - \Psi_{2}^{\rm SP})\rangle$ are all much stronger than $\langle{\rm cos}\ 2(\Psi_B - \Psi_{2})\rangle$ and $\langle(eB/m_{\pi}^{2})^{2}{\rm cos}\ 2(\Psi_B - \Psi_{2})\rangle$ for the two isobaric collisions. Moreover, deformation does affect the CME signals in isobaric collisions, especially for peripheral events in which the larger deformation leads to the weaker $\langle{\rm cos}\ 2(\Psi_B - \Psi_{2})\rangle$ and $\langle(eB/m_{\pi}^{2})^{2}{\rm cos}\ 2(\Psi_B - \Psi_{2})\rangle$. For case 1, the relative difference with respect to the spectator plane and that with respect to the participant plane look similar due to their small relative deformation difference. For case 2, the two relative differences look different due to their larger deformation difference. Since $\Psi_{2}^{\rm SP}$ has a much stronger correlation with $\Psi_B$ than $\Psi_{2}$, the $\Delta \gamma$ correlator with respect to $\Psi_{2}^{\rm SP}$ is expected to reflect much cleaner information about the CME signal due to different magnitudes of magnetic fields between two isobaric collisions with less influences of deformation.

\section{acknowledgments}
We thank X. G. Huang, Z. W. Lin, Q. Y. Shou, Z. Tu, and F. Q. Wang for their helpful discussions and comments. This work was supported by the National Natural Science Foundation of China under Grants No. 11890714, No. 11835002, 11421505, No. 11522547 and No. 11375251, the Key Research Program of the Chinese Academy of Sciences under Grant No. XDPB09. X. L. Zhao was supported by the Chinese Government Scholarship of Chinese Scholarship Council under CSC Grant No. 201804910796.


\begin{thebibliography}{99}

\bibitem{Kharzeev:2015znc}
  D.~E.~Kharzeev, J.~Liao, S.~A.~Voloshin and G.~Wang,
  Prog.\ Part.\ Nucl.\ Phys.\  {\bf 88}, 1 (2016)
  [arXiv:1511.04050 [hep-ph]].
\bibitem{Fukushima:2008xe}
  K.~Fukushima, D.~E.~Kharzeev and H.~J.~Warringa,
  Phys.\ Rev.\ D {\bf 78}, 074033 (2008)
  [arXiv:0808.3382 [hep-ph]].

\bibitem{Kharzeev:2004ey}
  D.~Kharzeev,
  Phys.\ Lett.\ B {\bf 633}, 260 (2006)
  [hep-ph/0406125].
\bibitem{Kharzeev:2007tn}
  D.~Kharzeev and A.~Zhitnitsky,
  Nucl.\ Phys.\ A {\bf 797}, 67 (2007)
  [arXiv:0706.1026 [hep-ph]].

\bibitem{Hattori:2016emy}
  K.~Hattori and X.~G.~Huang,
  Nucl.\ Sci.\ Tech.\  {\bf 28}, no. 2, 26 (2017)
  [arXiv:1609.00747 [nucl-th]].

\bibitem{Abelev:2009ac}
  B.~I.~Abelev {\it et al.} [STAR Collaboration],
  Phys.\ Rev.\ Lett.\  {\bf 103}, 251601 (2009)
  [arXiv:0909.1739 [nucl-ex]].
\bibitem{Abelev:2009ad}
  B.~I.~Abelev {\it et al.} [STAR Collaboration],
  Phys.\ Rev.\ C {\bf 81}, 054908 (2010)
  [arXiv:0909.1717 [nucl-ex]].
\bibitem{Abelev:2012pa}
  B.~Abelev {\it et al.} [ALICE Collaboration],
  Phys.\ Rev.\ Lett.\  {\bf 110}, no. 1, 012301 (2013)
  [arXiv:1207.0900 [nucl-ex]].
\bibitem{Bloczynski:2012en}
  J.~Bloczynski, X.~G.~Huang, X.~Zhang and J.~Liao,
  Phys.\ Lett.\ B {\bf 718}, 1529 (2013)
  [arXiv:1209.6594 [nucl-th]].
\bibitem{Deng:2016knn}
  W.~T.~Deng, X.~G.~Huang, G.~L.~Ma and G.~Wang,
  Phys.\ Rev.\ C {\bf 94}, 041901 (2016)
  [arXiv:1607.04697 [nucl-th]].

\bibitem{Bzdak:2010fd}
  A.~Bzdak, V.~Koch and J.~Liao,
  Phys.\ Rev.\ C {\bf 83}, 014905 (2011)
  [arXiv:1008.4919 [nucl-th]].
\bibitem{Schlichting:2010qia}
  S.~Schlichting and S.~Pratt,
  Phys.\ Rev.\ C {\bf 83}, 014913 (2011)
  [arXiv:1009.4283 [nucl-th]].
\bibitem{Wang:2009kd}
  F.~Wang,
  Phys.\ Rev.\ C {\bf 81}, 064902 (2010)
  [arXiv:0911.1482 [nucl-ex]].
\bibitem{Ma:2011uma}
  G.~L.~Ma and B.~Zhang,
  Phys.\ Lett.\ B {\bf 700}, 39 (2011)
  [arXiv:1101.1701 [nucl-th]].

\bibitem{Adamczyk:2013kcb}
  L.~Adamczyk {\it et al.} [STAR Collaboration],
  Phys.\ Rev.\ C {\bf 89}, no. 4, 044908 (2014)
  [arXiv:1303.0901 [nucl-ex]].
\bibitem{Wang:2016iov}
  F.~Wang and J.~Zhao,
  Phys.\ Rev.\ C {\bf 95}, no. 5, 051901 (2017)
  [arXiv:1608.06610 [nucl-th]].
\bibitem{Zhao:2018blc}
  J.~Zhao [STAR Collaboration],
  Nucl.\ Phys.\ A {\bf 982}, 535 (2019)
  [arXiv:1807.09925 [nucl-ex]].

\bibitem{Voloshin:2010ut}
  S.~A.~Voloshin,
  Phys.\ Rev.\ Lett.\  {\bf 105}, 172301 (2010)
  [arXiv:1006.1020 [nucl-th]].

 \bibitem{Huang:2017azw}
  X.~G.~Huang, W.~T.~Deng, G.~L.~Ma and G.~Wang,
  Nucl.\ Phys.\ A {\bf 967}, 736 (2017)
  [arXiv:1704.04382 [nucl-th]].
\bibitem{Deng:2018dut}
  W.~T.~Deng, X.~G.~Huang, G.~L.~Ma and G.~Wang,
  Phys.\ Rev.\ C {\bf 97}, no. 4, 044901 (2018)
  [arXiv:1802.02292 [nucl-th]].
\bibitem{Shi:2017ucn}
  S.~Shi, Y.~Jiang, E.~Lilleskov and J.~Liao,
  PoS CPOD {\bf 2017}, 021 (2018)
  [arXiv:1712.01386 [nucl-th]].
\bibitem{Magdy:2018lwk}
  N.~Magdy, S.~Shi, J.~Liao, P.~Liu and R.~A.~Lacey,
  arXiv:1803.02416 [nucl-ex].
\bibitem{Xu:2017zcn}
  H.~J.~Xu, X.~Wang, H.~Li, J.~Zhao, Z.~W.~Lin, C.~Shen and F.~Wang,
  Phys.\ Rev.\ Lett.\  {\bf 121}, no. 2, 022301 (2018)
  [arXiv:1710.03086 [nucl-th]].
\bibitem{Li:2018oec}
  H.~Li, H.~j.~Xu, J.~Zhao, Z.~W.~Lin, H.~Zhang, X.~Wang, C.~Shen and F.~Wang,
  arXiv:1808.06711 [nucl-th].
\bibitem{Sun:2018idn}
  Y.~Sun and C.~M.~Ko,
  Phys.\ Rev.\ C {\bf 98}, no. 1, 014911 (2018)
  [arXiv:1803.06043 [nucl-th]].

\bibitem{Voloshin:2018qsm}
  S.~A.~Voloshin,
  arXiv:1805.05300 [nucl-ex].
\bibitem{Lin:2004en}
  Z.~W.~Lin, C.~M.~Ko, B.~A.~Li, B.~Zhang and S.~Pal,
  Phys.\ Rev.\ C {\bf 72}, 064901 (2005)
  [nucl-th/0411110].
\bibitem{Wang:1991hta}
  X.~N.~Wang and M.~Gyulassy,
  Phys.\ Rev.\ D {\bf 44}, 3501 (1991).
\bibitem{Gyulassy:1994ew}
  M.~Gyulassy and X.~N.~Wang,
  Comput.\ Phys.\ Commun.\  {\bf 83}, 307 (1994)
  [nucl-th/9502021].
\bibitem{Zhang:1997ej}
  B.~Zhang,
  Comput.\ Phys.\ Commun.\  {\bf 109}, 193 (1998)
  [nucl-th/9709009].

\bibitem{Li:1995pra}
  B.~A.~Li and C.~M.~Ko,
  Phys.\ Rev.\ C {\bf 52}, 2037 (1995)
  [nucl-th/9505016].

\bibitem{Shou:2014eya}
  Q.~Y.~Shou, Y.~G.~Ma, P.~Sorensen, A.~H.~Tang, F.~Videb$\ae$k and H.~Wang,
  Phys.\ Lett.\ B {\bf 749}, 215 (2015)
  [arXiv:1409.8375 [nucl-th]].
\bibitem{Raman:1201zz}
  S.~Raman, C.~W.~G.~Nestor, Jr and P.~Tikkanen,
  Atom.\ Data Nucl.\ Data Tabl.\  {\bf 78}, 1 (2001).
\bibitem{Pritychenko:2013gwa}
  B.~Pritychenko, M.~Birch, B.~Singh and M.~Horoi,
  Atom.\ Data Nucl.\ Data Tabl.\  {\bf 107}, 1 (2016)
  Erratum: [Atom.\ Data Nucl.\ Data Tabl.\  {\bf 114}, 371 (2017)]
  [arXiv:1312.5975 [nucl-th]].
\bibitem{Moller:1993ed}
  P.~Moller, J.~R.~Nix, W.~D.~Myers and W.~J.~Swiatecki,
  Atom.\ Data Nucl.\ Data Tabl.\  {\bf 59}, 185 (1995)
  [nucl-th/9308022].
\bibitem{Bzdak:2011yy}
  A.~Bzdak and V.~Skokov,
  Phys.\ Lett.\ B {\bf 710}, 171 (2012)
  [arXiv:1111.1949 [hep-ph]].
\bibitem{Deng:2012pc}
  W.~T.~Deng and X.~G.~Huang,
  Phys.\ Rev.\ C {\bf 85}, 044907 (2012)
  [arXiv:1201.5108 [nucl-th]].
\bibitem{Zhao:2017rpf}
  X.~L.~Zhao, Y.~G.~Ma and G.~L.~Ma,
  Phys.\ Rev.\ C {\bf 97}, no. 2, 024910 (2018)
  [arXiv:1709.05962 [hep-ph]].

\bibitem{Deng:2017ljz}
  X.~G.~Deng and Y.~G.~Ma,
  Nucl.\ Sci.\ Tech.\  {\bf 28}, no. 12, 182 (2017).

\bibitem{Alver:2010gr}
  B.~Alver and G.~Roland,
  Phys.\ Rev.\ C {\bf 81}, 054905 (2010)
  Erratum: [Phys.\ Rev.\ C {\bf 82}, 039903 (2010)]
  [arXiv:1003.0194 [nucl-th]].
\bibitem{Ma:2010dv}
  G.~L.~Ma and X.~N.~Wang,
  Phys.\ Rev.\ Lett.\  {\bf 106}, 162301 (2011)
  [arXiv:1011.5249 [nucl-th]].
\bibitem{Chatterjee:2014sea}
  S.~Chatterjee and P.~Tribedy,
  Phys.\ Rev.\ C {\bf 92}, no. 1, 011902 (2015)
  [arXiv:1412.5103 [nucl-th]].
\bibitem{Abelev:2013cva}
  B.~Abelev {\it et al.} [ALICE Collaboration],
  Phys.\ Rev.\ Lett.\  {\bf 111}, no. 23, 232302 (2013)
  [arXiv:1306.4145 [nucl-ex]].

\bibitem{Adler:2001fq}
  C.~Adler, H.~Strobele, A.~Denisov, E.~Garcia, M.~Murray and S.~White,
  Nucl.\ Instrum.\ Meth.\ A {\bf 461}, 337 (2001).


\end{thebibliography}
\end{document}